\documentclass[twocolumn]{emulateapj}

\usepackage{amsmath}
\usepackage{url}
\usepackage{color}

\shortauthors{
Wong et al.
}

\shorttitle{
X-ray Observations of the Antlia Cluster
}
\slugcomment{Astrophysical Journal, submitted}

\begin{document}
\title{
{\it Suzaku} X-ray Observations of the Nearest Non-Cool Core Cluster, 
Antlia: Dynamically Young but with Remarkably Relaxed Outskirts
}

\author{
Ka-Wah Wong\altaffilmark{1,2},
Jimmy A. Irwin\altaffilmark{3},
Daniel R. Wik\altaffilmark{4,5},
Ming Sun\altaffilmark{6},
Craig L. Sarazin\altaffilmark{7},
Yutaka Fujita\altaffilmark{8},
and
Thomas H. Reiprich\altaffilmark{9},
}

\altaffiltext{1}{Eureka Scientific, Inc., 2452 Delmer Street Suite 100,
Oakland, CA 94602-3017, USA
}
\altaffiltext{2}{Department of Physics and Astronomy, Minnesota State 
University, Mankato, MN 56001, USA
}
\altaffiltext{3}{Department of Physics and Astronomy, University of 
Alabama, Box 870324, Tuscaloosa, AL 35487, USA
}
\altaffiltext{4}{Astrophysics Science Division, NASA/Goddard Space 
Flight Center, Greenbelt, MD 20771, USA
}
\altaffiltext{5}{The Johns Hopkins University, Homewood Campus, 
Baltimore, MD 21218, USA
}
\altaffiltext{6}{Physics Department, University of Alabama in 
Huntsville, Huntsville, AL 35899, USA
}
\altaffiltext{7}{Department of Astronomy, University of Virginia, P.O. 
Box 400325, Charlottesville, VA 22904-4325, USA
}
\altaffiltext{8}{Department of Earth and Space Science, Graduate School 
of Science, Osaka University, Toyonaka, Osaka 560-0043, Japan
}
\altaffiltext{9}{Argelander Institut f\"{u}r Astronomie, Universit\"{a}t 
Bonn, Auf dem H\"{u}gel 71, D-53121, Germany
}
\email{kw6k@email.virginia.edu}

\begin{abstract}
We present the results of seven {\it Suzaku} mosaic observations ($>$200\,ks 
in total) of the nearest non-cool core cluster, the Antlia Cluster (or 
Group), beyond its degree-scale virial radius in its 
eastern direction.
The temperature is consistent with the scaled profiles of 
many other clusters. Its pressure follows the universal profile.  The 
density slope in its outskirts is significantly steeper than that of the 
nearest cool core cluster (Virgo) with a similar temperature as Antlia, 
but shallower than those of the massive clusters.  The entropy increases 
all the way out to $R_{200}$, which is consistent in value with the baseline model 
predicted by a gravity heating-only mechanism in the outskirts.  
Antlia is quite relaxed in this direction.
However, the entropy inside $\sim$$R_{500}$ is significantly higher than 
the baseline model, similar to many other nearby low mass clusters or 
groups.  The enclosed gas-mass fraction does not exceed the cosmic value 
out to $1.3 R_{200}$.  Thus, there is no evidence of significant gas 
clumping, electron-ion non-equipartition, or departure from the 
hydrostatic equilibrium approximation that are suggested to explain the 
entropy and gas fraction anomalies found in the outskirts of some massive 
clusters. We also present scaling relations for the gas fraction 
($f_{\rm gas,200}$), entropy ($K_{200}$), and temperature ($T_{500}$) 
using 22 groups and clusters with published data in the literature. The 
enclosed baryon fraction at $R_{200}$ is 
broadly consistent with the cosmic 
value.  The power law slope of the $K_{200}$--$T_{500}$ relation is 
$0.638 \pm 0.205$.  The entropy deficit at $R_{200}$ cannot be fully 
accounted for by the bias or deviation in the gas fraction.

\end{abstract}

\keywords{
X-rays: galaxies: clusters ---
galaxies: clusters: intracluster medium ---
galaxies: groups: individual (the Antlia Cluster) ---
intergalactic medium ---
cosmology: large-scale structure of universe
}

\section{Introduction}
\label{sec:intro}

X-ray observations have shown that the study of the intracluster medium 
(ICM) can be used to test plasma physics under extreme conditions 
that cannot be achieved in terrestrial laboratories, as well as an 
important cosmological probe.
However, the study of cosmology using clusters of galaxies relies 
heavily on the understanding of cluster physics. For precision 
cosmology, systematic uncertainties at even the percent level are quite 
significant.  Even though present X-ray measurements of clusters for 
cosmological studies are limited to the inner regions ($\lesssim 
R_{500}$ due to sensitivity limits\footnote{$R_{\Delta}$ 
is the radius within which the mean total 
mass density of the cluster is $\Delta$ times the critical density of 
the universe.  The virial radius $R_{\rm vir}$ is defined as a radius 
within which the cluster is virialized.  For the Einstein-de Sitter 
universe, $ R_{\rm vir} \approx R_{200}$, while for the standard 
$\Lambda$CDM Universe, $ R_{\rm vir} \approx R_{100}$.  In this paper,
we also call $R_{200}$ the virial radius since this definition 
is still widely 
used in the literature.}), many X-ray 
observations and Sunyaev-Zel'dovich (SZ) observations have 
already extended to larger radii.  Thus, a full understanding of the 
outer parts of clusters is essential.

Compared to the central regions, cluster envelopes were thought to 
be relatively simple because they are less subject to additional physics 
including cooling, and AGN feedback, and that the cluster outer 
regions might provide better cosmological probes.  This appeared to be 
supported by earlier observational results with {\it ROSAT} that the 
X-ray surface brightness around $R_{200}$ is generally consistent with 
numerical simulations.  However, more recent observations, primarily 
with {\it Suzaku}, have shown a number of unexpected results \citep[see 
a recent review by][]{RBE+13}. For example, it has been suggested that 
the ICM near $R_{200}$ can be rather clumpy, which means the enclosed 
gas-mass fraction measured with X-ray can be biased high and may exceed the 
cosmic baryon fraction \citep[e.g.,][]{NL11,Sim+11};
although \citet{WFS+12} and \citet{Oka+14} pointed out that clumpiness in 
the simulations by \citet{NL11} becomes dominant only beyond $R_{200}$.
The entropy near 
$R_{200}$ of some massive clusters measured with {\it Suzaku} appears to 
be significantly lower than the predictions from the gravity 
heating-only models \citep[see, e.g.,][and references therein]{WFS+13}, 
although the {\it ROSAT} results are less significant \citep{EMV+13}. 
This low entropy may indicate that the collisional equilibrium of electrons 
and ions has not been achieved in the low density hot gas in cluster 
outskirts \citep[e.g.,][]{WS09,Hos+10,AHI+11,WSJ11,ANL+15}, which may 
introduce biases in cosmological parameters \citep{WSW10}. Pressure 
support from turbulence/bulk motions \citep[e.g.,][]{LKN09} or cosmic 
rays \citep{FOY13} are also thought to be important around $R_{200}$, 
resulting in the breakdown of the hydrostatic equilibrium (HSE) 
approximation if only the thermal pressure is considered 
\citep[e.g.,][]{Kaw+10,Ich+13,Oka+14}.  More than a dozen clusters have 
been studied with spatially and spectrally resolved gas profiles out to 
$\sim$$R_{200}$ to understand these anomalous effects \citep[see again 
the review by][]{RBE+13}.

Mapping the nearest clusters with mosaic pointings to cover angular 
scales of order degrees has revealed the complex ICM structures in 
tremendous spatial detail.  A direct comparison to the nearest clusters 
with different physical properties in our neighborhood is certainly 
important to understand the cluster outskirts. For example, the nearby 
massive cool core cluster, Perseus \citep[distance $D$\,=\,70\,Mpc, 
$R_{200}$\,=\,1.8\,Mpc\,=\,1.4\arcdeg:][]{Sim+11,Urb+14}, 
shows significant bias in 
the gas fraction and entropy deficit near $R_{200}$, as well as strong 
azimuthal variations.  The massive merging cluster, 
Coma \citep[$D$\,=\,100\,Mpc, $R_{200}$\,$\sim$\,2\,Mpc\,=\,1.2\arcdeg:][]{AHI+13,Sim+13}, shows both 
a dynamical active environment with a merging group in one direction and 
a relaxed environment along the other directions.

The nearest cluster (or group), Virgo 
($D$\,=\,16\,Mpc, $R_{200}$\,=\,1\,Mpc\,=\,4\arcdeg), 
which is also a cool core with a lower mass 
($M_{500}$\,$\approx$\,$10^{14}M_{\odot}$) 
and temperature ($\approx 2.3$\,keV) compared 
to the massive clusters ($M_{500}\gtrsim 5 \times 10^{14}M_{\odot}$) 
mentioned above, has also been mapped out to $R_{200}$ 
\citep{UWS+11,SWU+15}.  Low mass galaxy groups are in fact also very 
important to cosmological studies.  Because of their high abundance, 
they contribute significantly to the SZ power spectrum at angular scales 
of $l \approx 3000$ that are sensitive to cosmological parameters 
\citep[][]{TBO11}. The previous {\it XMM-Newton} result for Virgo 
suggests that its density profile near $R_{200}$ has a power law
($n_e \propto r^{-\alpha}$) slope 
of $\alpha = 1.2\pm0.2,\footnote{The errors for Virgo and Perseus have 
been converted to 90\% confidence.}$ which is significantly flatter than 
for massive cluster counterparts (e.g., $\alpha = 1.7\pm0.2$ for 
Perseus, \citealt{Urb+14}; $\alpha = 2.27\pm0.07$ for A1795, 
\citealt{Bau+09}; and $\alpha =2$--3 for a stacked {\it Chandra} sample, 
\citealt{MSF+15}). Therefore, it is natural to wonder whether non-cool 
core groups behave in the same manner or differently, compared with massive 
clusters or the Virgo cool core group; understanding this will have 
deep implications for both cluster studies and cosmology.

\begin{figure*}
\includegraphics[width=.48\textwidth, angle=0]{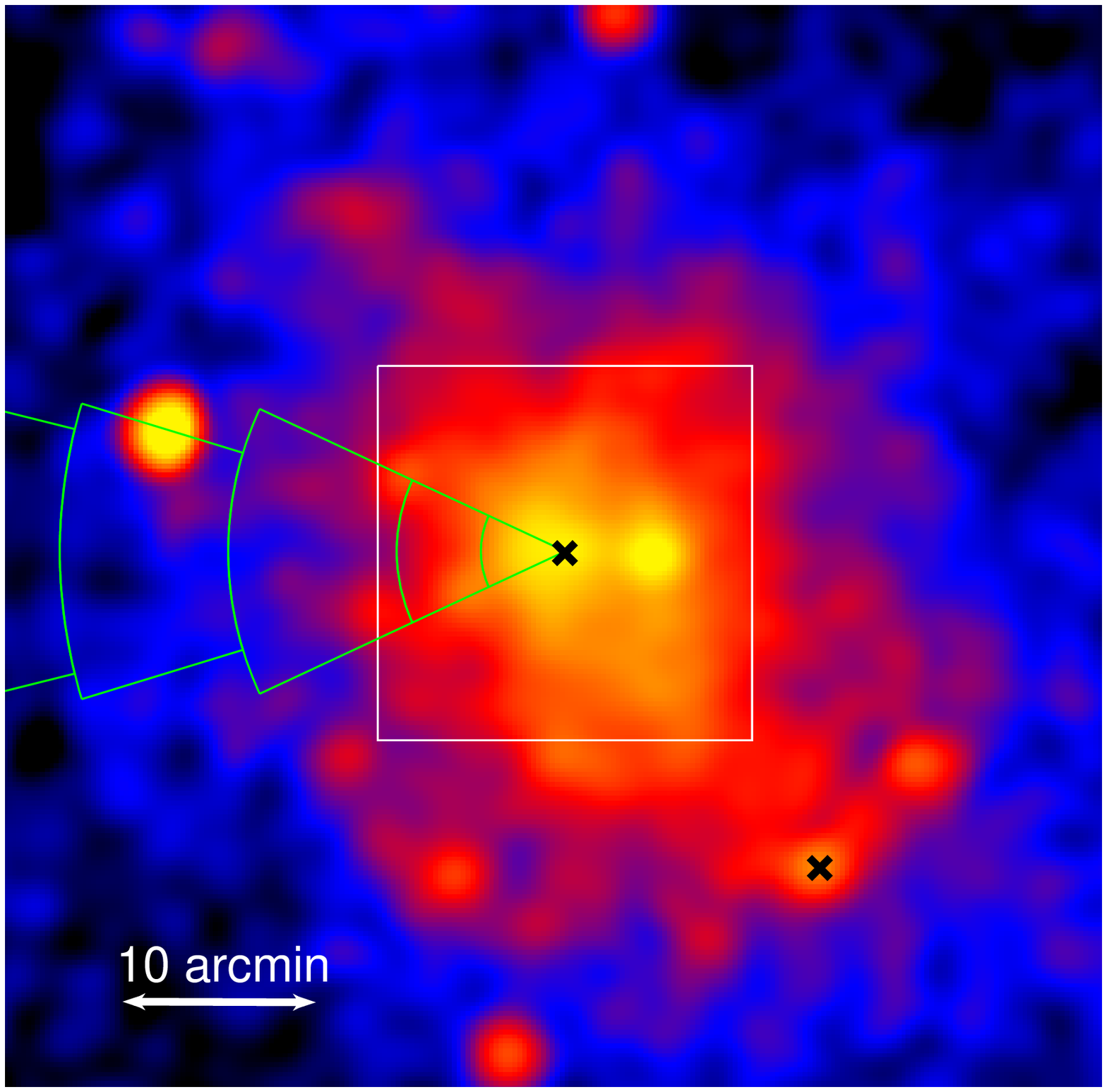}
\includegraphics[width=.48\textwidth, angle=0]{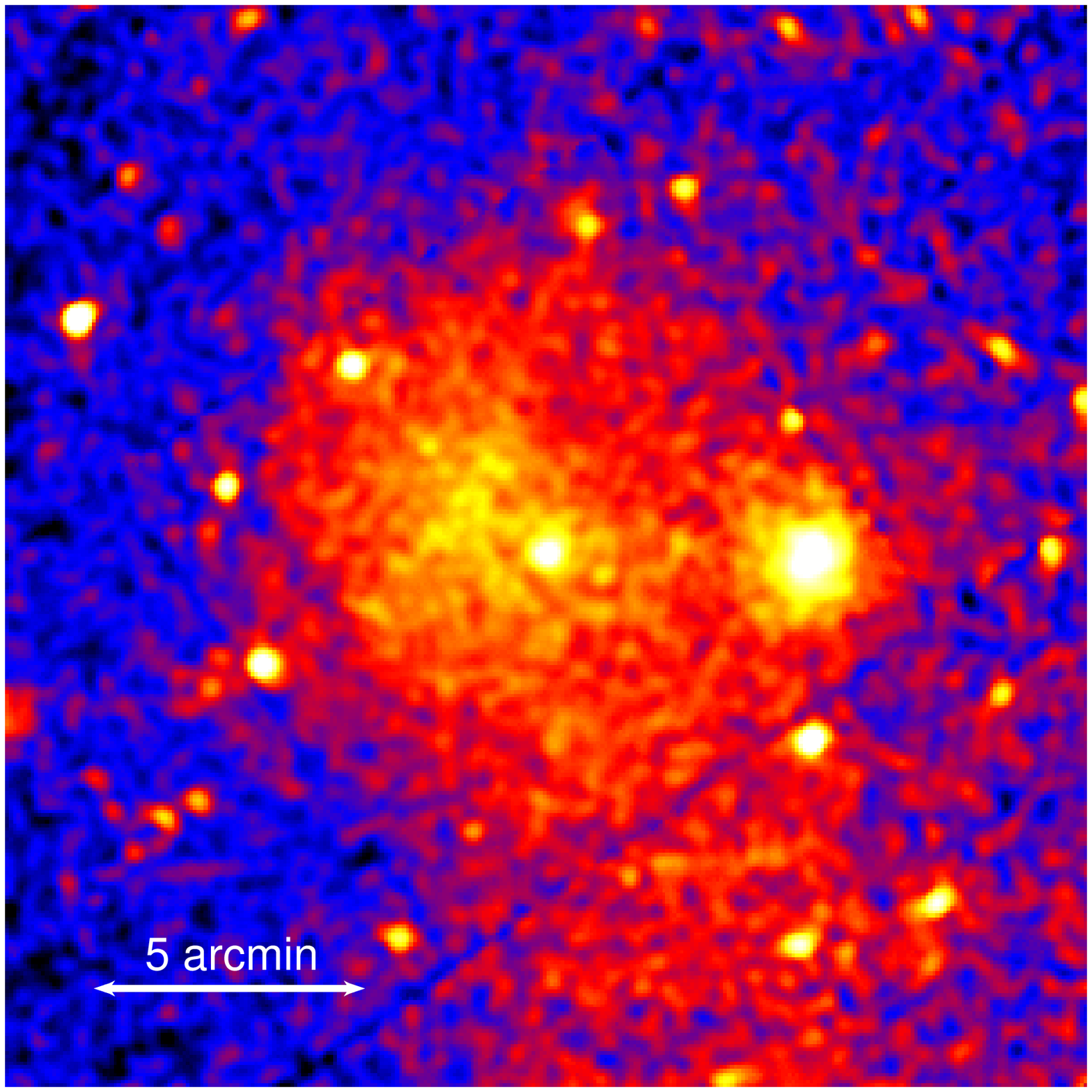}
\includegraphics[width=1.\textwidth, angle=0]{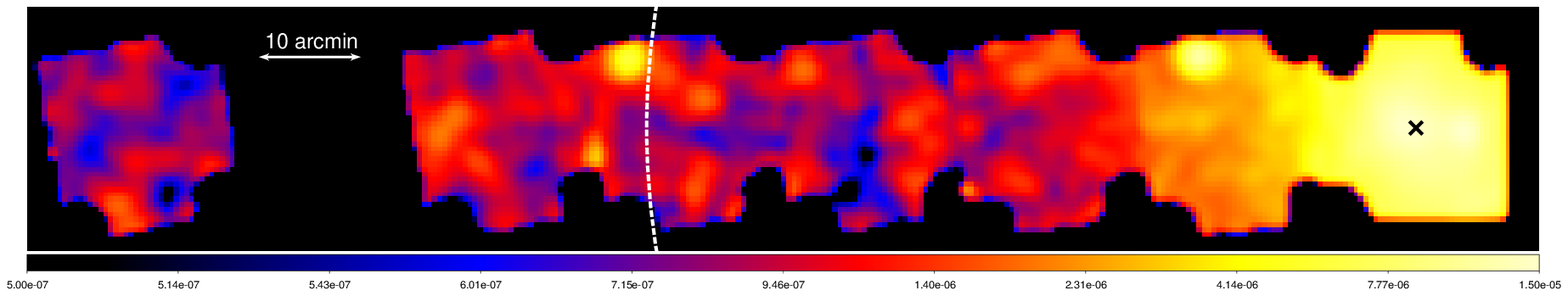}
\includegraphics[width=1.\textwidth, angle=0]{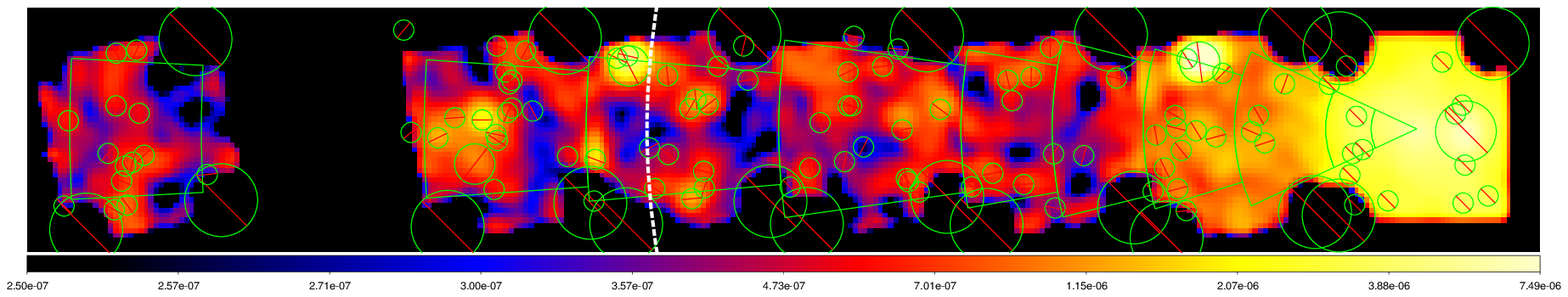}
\caption{Smoothed {\it ROSAT} PSPC image of the central regions of the Antlia
Cluster in the 0.5--2\,keV band is shown in the upper left.
North is up, while east is left.  
The color represents the X-ray intensity from high (yellow--red) to low 
(dark blue).
The upper and lower black crosses indicate the locations of NGC~3268 and 
NGC~3258, respectively.
The X-ray peak about 5\arcmin\ west of NGC~3268 is probably a background 
cluster (Section~\ref{sec:ptsrc}).
The innermost four spectral extraction areas of the {\it Suzaku} 
data analysis are shown in green pie regions.
The white square indicates the field of view of the smoothed 
{\it XMM-Newton} MOS2 image (Obs. ID: 0691950201: PI: E.~T. Million) 
in the 0.5--2\,keV band shown on the upper right panel.
Also shown are the exposure corrected, background subtracted, 
and smoothed soft-band 
(0.6--2.0\,keV: middle) and hard-band (2.0--7.0\,keV: bottom) {\it Suzaku} 
images of Antlia with an image pixel binning size = 0\farcm5. Both 
images were smoothed with a Gaussian kernel of FWHM $\approx$ 1\arcmin.  
The color bars show the surface brightness in units of 
count\,s$^{-1}$\,cm$^{-2}$\,arcmin$^{-2}$.
The seven {\it Suzaku} pointings from the right to the left are E0, E1,  
E2, E3, E4, E5, and the local background field EB.
The dashed white arc on the middle image has a radius of 
$R_{200} = 887$~kpc~$ = 76\arcmin$
centered on the peak of the extended emission (black cross). The removed point 
sources (or compact structures) and the calibration sources (at field corners) 
are shown in solid green circles (with a red line across) on the bottom 
image.  The spectral extraction areas are shown in solid green pie 
regions on the bottom image. The unresolved diffuse emission in the 
soft-band image is dominated by hot gas within $\lesssim 30\arcmin$ while the 
hard-band image is dominated by CXB beyond $\gtrsim 20\arcmin$.
}
\label{fig:image}
\vskip 10mm
\end{figure*}

Sitting at a distance of $D$\,=\,39.8\,Mpc \citep{CBR+05}, the Antlia Cluster 
is the nearest non-cool core cluster (or a group with similar size to 
the Virgo Cluster) and also the third closest galaxy cluster after the 
Fornax cool core cluster (group).  X-ray observations suggest that the 
Antlia core is approximately isothermal ($kT \sim 2$\,keV) with no 
significant excess central brightness \citep{NMF+00}.  The X-ray 
emission is centered on the bright elliptical galaxy NGC 3268 and
is elongated toward the southwest where there is a subgroup centered on 
another bright elliptical galaxy NGC 3258, indicating that the Antlia 
Cluster is accreting along this direction \citep[Figure~\ref{fig:image}; see 
also][]{PYS97,NMF+00}.  Optical observations suggest that Antlia is 
dynamically younger than Virgo and Fornax,
and the large-scale filament structure of the galaxies are along 
the northwest and southeast directions.
\citep{FS90,SBR+08,HJC+15}.

Here, we report the results from mosaic {\it Suzaku} observations of Antlia 
to the east direction out to $1.3 R_{200}$.  Compared to studying cool 
core clusters/groups, the non-cool core Antlia is less subject to 
systematic uncertainty due to the telescope scattering of X-ray photons 
from the bright center to the outer regions, which are of greatest
interest here.  The direction was chosen so that it is away from the
accreting
filament
direction where substructures induced by accretion can disturb 
the global properties significantly.  
Thus, the eastern direction should be more relaxed compared to the 
filament directions, as we will also show in Section~\ref{sec:dis1} 
below.
It was also chosen to avoid many of 
the point sources and the high background seen on the {\it ROSAT} image.  We 
made use of the {\it Chandra} and {\it XMM-Newton} data to minimize the 
uncertainty due to point sources unresolved by {\it Suzaku} 
\citep{MBG+12}.  We also compared our results with {\it ROSAT} data to 
ensure that the soft X-ray emission determined with {\it Suzaku} was 
robust \citep[see, e.g.,][]{EMG+11}.

With the growing number of clusters and groups measured out 
to $R_{200}$, it is possible to study the scaling relations between the 
gas fraction ($f_{\rm gas,200}$), entropy ($K_{200}$), and temperature
where the gas fraction and entropy are of great interest to constrain 
cosmological parameters and to understand the thermodynamic history of 
cluster or group formations, respectively \citep[see, 
e.g.,][]{SVD+09,Pra+10}.  Thus, we compiled the observational data of 22 
groups and clusters from the literature to study the $f_{\rm 
gas,200}$--$T_{500}$ and $K_{200}$--$T_{500}$ relations out to 
$R_{200}$.

The paper is organized as follows.  Section~\ref{sec:obs} describes the 
X-ray observations and data analysis. 
The following sections describe the observational results, including
the surface brightness (Section~\ref{sec:SB}); 
temperature (Section~\ref{sec:tprofile}); density, 
pressure, and entropy (Section~\ref{sec:DPE}); 
mass (Section~\ref{sec:mass}); and equilibration 
timescale (Section~\ref{sec:time}) profiles of the hot gas.
We present the $f_{\rm gas,200}$--$T_{500}$ and 
$K_{200}$--$T_{500}$ relations in Section~\ref{sec:scaling}.
We compare Antlia to other galaxy groups and massive clusters and 
discuss the implications in Section~\ref{sec:discussions}.
We 
summarize our conclusions in Section~\ref{sec:summary}.  
Appendix~\ref{sec:EM} presents the density deprojection methods.
Systematic 
uncertainties in spectral modeling are addressed in detail in 
Appendix~\ref{sec:app1}.

At a distance of 39.8\,Mpc, the angular scale of the Antlia Cluster is 
11.6~kpc/1\arcmin. We assume $H_0$\,=\,70\,km\,s$^{-1}$\,Mpc$^{-1}$, 
$\Omega_M$\,=\,0.3, and $\Omega_{\Lambda}$\,=\,0.7. The average X-ray 
temperature of Antlia beyond the core region is determined to be 
$T_X$\,=\,1.54\,keV (Section~\ref{sec:tprofile}). 
We adopted the scale radius of 
$R_{500}$\,=\,591\,$h_{70}^{-1}{\rm \,kpc}\, (T_X/1.54{\rm ~keV})^{0.55}$\,=\,51\arcmin\ inferred from the X-ray scaling relation of \citet{SVD+09}.  
For low mass galaxy groups, the concentration parameter $c_{200}$ of the 
NFW dark matter model is about six, and therefore $R_{200} \approx 1.5 
R_{500} = 887{\rm ~kpc} = 76\arcmin$ \citep[e.g., Figure~4 in][]{YRS09}. 
Errors are given at 90\% confidence level in this paper unless otherwise 
specified.

\section{X-ray Observations and Spectral Analysis}
\label{sec:obs}

\subsection{Data Reduction}
\label{sec:reduction}

The Antlia Cluster (Figure~\ref{fig:image}) was observed with {\it 
Suzaku} at the center in 2008 (PI: T. Kitaguchi), and also along the east 
direction for six pointings out to a radius of $\sim$135\arcmin\ in 2012 (PI: 
K.-W. Wong).  The central pointing (E0) was observed for 66\,ks.  Each of 
the next three pointings from the center (E1--E3) were observed for 
23--26\,ks, and the outermost three pointings (E4, E5, and EB) were each 
observed for 46--47\,ks. These seven observations are listed in 
Table~\ref{table:obs}.

The {\it Suzaku} XIS0, XIS1, and XIS3 data were reduced using the {\tt 
HEAsoft} package version 6.15\footnote{http://heasarc.gsfc.nasa.gov/docs/software/lheasoft/} and the {\it 
Suzaku} {\tt CALDB} version 20140203.  All the data were reprocessed 
using the FTOOLS {\tt aepipeline} script in the {\tt HEAsoft} package.  
All the standard screening criteria were applied.\footnote{http://heasarc.nasa.gov/docs/suzaku/processing/criteria\_xis.html}  In 
addition, we selected data with the geomagnetic cut-off rigidity COR2 
$>$ 6~GV to reduce the particle-induced non-X-ray background (NXB), and 
removed events in regions illuminated by the XIS calibration sources. 
For the XIS1 data observed in 2012, the increase of charge injection 
enhanced the NXB.  To reduce the NXB, we have also followed the 
recommended procedures\footnote{http://heasarc.gsfc.nasa.gov/docs/suzaku/analysis/abc/\\node8.html} in the 
pipeline processing.  The data with $3\times3$ and $5\times5$ editing 
modes for each pointing were merged.  
We have examined the 12--14\,keV light curves of the screened data and did 
not find any flares in the NXB.

For each pointing, we constructed light curves and removed data with 
count rates that deviated by $>$3$\sigma$ from the mean to avoid 
potential flares due to solar wind charge exchange (SWCX) or 
astrophysical sources, among others.  Two energy bands of 0.5--8.0 and 0.4--1.0\,keV 
were used, with the latter energy band chosen to check for contamination 
due to SWCX.  After that, we visually inspected each light curve and 
found that in the observation E4, the count rate of the second half  
of the observation in 0.4--1.0\,keV was about twice as high as the first 
half of the observation.  
The enhanced solar proton flux by up to a factor of 
seven measured by 
the {\it Advanced Composition Explorer (ACE)} during this observation 
suggests an increase in SWCX, and therefore 
we have also removed the second half of this 
observation.
The 
cleaned effective exposure times are listed in Table~\ref{table:obs}.

NXB files for each detector and observation were created according to 
the cut-off rigidity weighting using the FTOOLS {\tt xisnxbgen} script 
with the same filtering criteria mentioned above.  We selected NXB data 
within\,$\pm$300 days of each observation.  These backgrounds were 
subtracted from the image and spectral analysis below.

\begin{deluxetable}{ccccc}
\tabletypesize{\scriptsize}
\tablewidth{0pt}
\tablecolumns{5}
\tablecaption{Observations
\label{table:obs}
}
\tablehead{
\colhead{Name} &
\colhead{Obs. ID} &
\colhead{Obs. Date} &
\colhead{Exp.(ks)$^{\rm a}$} &
\colhead{$n_{\rm H}$$^{\rm b}$}
}
\startdata
{\it Suzaku} & & & & \\
\hline
Antlia E0 & 802035010 & 2007 Nov 19 & 55 & 8.84 \\
Antlia E1 & 807066010 & 2012 Jun 13 & 20 & 8.90 \\
Antlia E2 & 807067010 & 2012 Jun 14 & 21 & 9.00 \\
Antlia E3 & 807068010 & 2012 Jun 15 & 19 & 9.07 \\
Antlia E4 & 807069010 & 2012 Jun 16 & 17 & 9.14 \\
Antlia E5 & 807070010 & 2012 Jun 17 & 39 & 9.09 \\
Antlia EB & 807071010 & 2012 Jun 18 & 38 & 8.17 \\
\hline
\\
{\it XMM-Newton} & & & & \\
\hline
Antlia E0 & 0604890101 & 2010 Jan 03 & 52/48 & \\
\hline
\\
{\it Chandra} & & & & \\
\hline
Antlia E1 & 15090 & 2013 Nov 20 & 7 & \\
Antlia E2 & 15089 & 2013 Nov 22 & 7 & \\
Antlia E3 & 15088 & 2013 Jul 02 & 7 & \\
Antlia E4 & 15086 & 2013 Nov 04 & 7 & \\
Antlia E5 & 15085 & 2013 Apr 05 & 7 & \\
Antlia EB & 15087 & 2013 Nov 04 & 7 &
\enddata
\tablenotetext{a}{Effective exposure time after cleaning.  The {\it 
XMM-Newton} exposures are for MOS/PN.}
\tablenotetext{b}{Hydrogen column density in units of $10^{20}$\,cm$^{-2}$ 
\citep{WSB+13}.}
\end{deluxetable}

\subsection{Point Source Removal}
\label{sec:ptsrc}

The bottom panel of Figure~\ref{fig:image} shows the hard-band 
(2--7\,keV) {\it Suzaku} image, with green circles indicating the XIS calibration 
sources and other contamination sources removed during the analysis.  

The contamination sources were detected as follows:
The central region (E0) of the Antlia was observed with {\it XMM-Newton} 
in 2010 (ObsID 0604890101, PI: M. Machacek, Table~\ref{table:obs}).  
Because the
ICM emission from the central region of the Antlia Cluster is 
bright enough and background contamination from point sources or compact 
structures is not important, we simply took the archival {\it 
XMM-Newton} full EPIC image in 0.2--12\,keV and detected sources using 
the {\tt CIAO wavdetect}.  
We examined the brightest 20 sources, where the flux is above the {\it 
Chandra} detection limit of $\approx$$1.4 \times 
10^{-14}$\,erg\,s$^{-1}$\,cm$^{-2}$ in 2--10\,keV (see below), for potential 
source removal in the {\it Suzaku} data analysis.
One of them is located at the center of 
the Antlia Cluster and is the peak of the ICM emission; therefore, it 
should not be excluded from the data analysis.  Four other sources are 
outside the {\it Suzaku} field of view (FOV).  One source 
(10:29:36.725, -35:19:36.38) detected is extended and is about 
$5\arcmin$ away from the center of the Antlia Cluster to the west 
direction.   
Spectral analysis indicates that the emission is thermal.  The location 
of its Fe K line suggests that
it is probably a background cluster located at a high 
redshift of $z$\,=\,0.4.  It has a temperature of 4\,keV and 
$R_{500}$\,$\approx$\,0.9\,Mpc\,$\sqrt{T_X/4\,{\rm keV}}$\,$\approx$\,3\arcmin.
A circular region of 3\arcmin\ in the radius of 
this background cluster candidate is excluded from the {\it Suzaku} data 
analysis.  We exclude all other sources using circular regions of 
1\arcmin\ in radius.  We visually inspected the {\it Suzaku} E0 data and 
did not find any source structure that was not detected by the {\it XMM-Newton} 
observations.

The fainter outer regions, especially near the virial radius, 
could be more 
subject to point source contaminations.  
We obtained six {\it Chandra} ACIS-I observations in 2013 
(PI: K.-W. Wong, Table~\ref{table:obs}), centered on each 
of the six {\it Suzaku} pointings away from the cluster center.
Each {\it Chandra} pointing was observed for 7\,ks. The {\it Chandra} 
ACIS-I observations cover most of the {\it Suzaku} XIS observed regions 
due to their similar FOVs. We created images with 0\farcs492 pixels 
in a broad energy band (0.1--10\,keV), and detected point sources with 
{\tt CIAO wavdetect}.  Sources detected with $>$3$\sigma$ were removed 
during the {\it Suzaku} data analysis.  
The detection limit is about $1.4 \times 
10^{-14}$\,erg\,cm$^{-2}$\,s$^{-1}$ in 2--10\,keV.
We also visually inspected {\it 
Suzaku} images and excluded obvious bright point-like sources missed by 
the {\tt CIAO wavdetect} script.  Most of these sources were removed 
using a circular region size of $1\arcmin$ in radius, with a few 
exceptions where larger sizes (1\farcm5 or 2\arcmin) were used for those with
emission clearly extending larger than $1\arcmin$ in radius.

The half-power diameter (HPD) of {\it Suzaku} is about 2\arcmin\, and 
therefore about half the photons from point sources fell outside our 
point source removal regions, and some of them entered the spectral 
extraction regions as residual signals.  We took into account the 
residual contributions in the spectral analysis (Section~\ref{sec:bg}).
We note that the {\it XMM-Newton} or {\it Chandra}
observations were taken close in time, but not simultaneously with 
the {\it Suzaku} observations.
Because these residual signals make a relatively small contribution to 
the spectra 
(Section~\ref{sec:SB}), the potential time variation of the 
integrated flux of the point sources should not be important.

\subsection{Spectral Analysis}
\label{sec:spec}

For each {\it Suzaku} XIS detector (XIS0, XIS1, and XIS3), we extracted 
spectra in pie regions centered on the central peak of the extended 
X-ray emission.  These pie regions, as well as the contamination 
exclusion regions, are shown in the lower panel of 
Figure~\ref{fig:image}.  The redistribution functions (RMF files) for 
each spectrum were generated using the FTOOLS {\tt xisrmfgen} script.  
The corresponding ancillary response functions (ARF files) were 
generated using the FTOOLS {\tt xissimarfgen} script.  
Because the spatial distributions of the ICM and background are 
different, different ARF files are required for spectral fitting.

When generating the ARF files for the ICM component for regions 
within 36\arcmin, we used the X-ray surface profile of the Antlia Cluster as 
the input for the {\tt xissimarfgen}.  For the inner 9\arcmin, the 
surface profile was modeled as a $\beta$-model fitted to the {\it 
XMM-Newton} data.  
The region beyond 9\arcmin\ was modeled using
the {\it Suzaku} data with the following iterative 
procedure.
We first generated the ARF files using the standard 20\arcmin\ radius 
uniform surface brightness source.  We then determined the
``initial'' ICM emission measure
using these uniform ARF files (using the same fitting procedure as the final 
spectral analysis below).  
We then fitted a double $\beta$-model to the ``initial'' ICM emission 
measure profile.  This double $\beta$-model was used to calculate 
the ``initial'' surface brightness profile for the proper ARF files generation.
We found that 
the final ICM emission measure determined using the proper ARF files
is generally different from this ``initial'' emission measure by
$<$10\% and smaller than the emission profile error, justifying the 
iterative procedure
\citep[see also][]{Bau+09}.
For regions beyond $\sim$$20\arcmin$--$30\arcmin$, the systematic 
uncertainty in the stray light calibration is very large, and therefore 
the ARF files generated using the cluster model input may not be 
reliable (E. D. Miller 2015, private communication; see also 
Section~6.2.6 of the {\it Suzaku} Technical Description\footnote{http://heasarc.gsfc.nasa.gov/docs/suzaku/prop\_tools/suzaku\_td/
suzaku\_td.html}).
Using {\tt XISSIM}, we found that the cluster stray light beyond 
$\sim$36\arcmin\ is less than $\sim$6\% of the NXB.
Therefore, the standard 20\arcmin\ radius uniform surface brightness 
source was used as the input for the {\tt xissimarfgen} for regions beyond 
36\arcmin.  
We checked that the differences of best-fit hot
gas parameters are less than 2\% when 
using the ARF files generated by the two methods for the 
regions between 27\arcmin\ and 36\arcmin.
The difference should be 
negligible for regions beyond that, justifying the use of uniform 
surface brightness as the input.  
For the background model fitting, the standard uniform ARF files were 
used in all regions.

As mentioned in 
Section~\ref{sec:reduction}, NXB spectra with proper cut-off-rigidity 
weighting were extracted from night-earth data using the FTOOLS {\tt 
xisnxbgen}.  For the six observations taken in 2012 (E1--E5 and EB), we 
followed the standard procedures in the {\it Suzaku} Data Reduction 
Guide
to mitigate the increase in NXB for XIS1 due to the 6\,keV charge 
injection.\footnote{http://heasarc.gsfc.nasa.gov/docs/suzaku/analysis/abc/abc.html}
Because of the similar responses, the XIS0 and XIS3 (FI) 
source and background spectra and the response files were 
combined.  We analyzed spectra between 0.6 and 7.0\,keV.  The lower 
energy limit was chosen to minimize the {\it Suzaku} calibration 
uncertainties below $\sim$0.6\,keV, while the upper energy limit is 
chosen due to the dominating NXB above $\sim$7\,keV.  All the spectra 
were grouped with a minimum of one count per bin, and were first fitted 
using the c-statistic in the X-ray Spectral Fitting 
Package
({\tt XSPEC}).\footnote{http://heasarc.nasa.gov/xanadu/xspec/} 
We then assessed the best-fit parameters (as median) and uncertainties 
using the MCMC method in {\tt XSPEC}.

\subsubsection{ICM Spectral Model}
\label{sec:ICM}

The ICM emission was modeled as an absorbed, optically thin thermal 
plasma model {\tt PHABS*APEC} using the atomic database {\tt AtomDB} 
version 2.0.2 \citep{SBL+01,FJS+12}.  For each pointing, 
the absorption was fixed to the Galactic value (Table~\ref{table:obs}) 
determined by \citet{WSB+13}. These values are generally $\sim$25--30\% 
higher than those determined by \citet{KBH+05}.  We have included this 
uncertainty by varying the Galactic values by $\pm30\%$ in our analysis 
(Appendix~\ref{sec:app1}). 
The redshift is fixed at $z$\,=\,0.00933 
(NASA/IPAC Extragalactic Database: NED).  
We adopted the solar abundance table from \citet[][hereafter {\tt 
aspl}]{AGS+09} for the ICM model, as well as the background model below.  
We have also assessed the abundance tables from 
\citet[][hereafter {\tt
angr}]{AG89}, \citet[][hereafter {\tt grsa}]{GS98}, and 
\citet[][hereafter {\tt lodd}]{Lod03}, which are widely used in the 
literature.  These systematic uncertainties have been taken into account 
in our analysis (Appendix~\ref{sec:app1}).
The metallicity was thawed for 
the inner regions within 18\arcmin.  For regions beyond that, where 
metallicity cannot be constrained, we fixed it to the lowest value of 
0.15\,$Z_{\odot}$ obtained in the inner region as our nominal model.  We 
have also fixed the metallicity to the maximum value of 0.4\,$Z_{\odot}$ 
determined at the center, and included this systematic uncertainty in our 
analysis (Appendix~\ref{sec:app1}).

We carried out both projected and deprojected spectral analysis for the 
ICM component.  For the projected spectral analysis, a single {\tt 
PHABS*APEC} model was used for the ICM component for each spectral 
region.  For the deprojected spectral analysis, the mixed model of {\tt 
PROJCT*(PHABS*APEC)} was used instead.  We assume the ICM to be 
spherically symmetric for the {\tt PROJCT} model, and the optional 
keywords for position angles have been properly adjusted to account for 
the partial annular spectral regions.

\subsubsection{Background Model}
\label{sec:bg}

The outermost pointing beyond the virial radius (EB) was chosen to be a 
local background.  The NXB is about 20 (35)\% of the total counts for 
either the XIS FI and BI detectors in the 0.6--2.0 (0.6--7.0)\,keV band 
in the spectral extraction region of this pointing; it is subtracted 
(from all the other data as well) during the analysis.  Varying the NXB 
by $\pm 5\%$ \citep[90\% confidence of systematic 
uncertainty;][]{Taw+08} does not change the results of the paper 
qualitatively (Appendix~\ref{sec:app1}).  This systematic error has been 
included in our data results.

The residual X-ray background is primarily the 
residual signals from the removed point sources, the remaining 
unresolved cosmic X-ray background (CXB),
Galactic X-ray foreground (GXB), and 
potential contamination of line emission from SWCX.  

To take into account the residual signals from the removed point 
sources, we first fitted the {\it Chandra} and {\it XMM-Newton} spectra 
for the removed point sources using an absorbed power law model with 
all the parameters thawed.  Using the {\tt xissim} script, we simulated 
{\it Suzaku} event files of these removed point sources for each 
pointing based on the spectra and fluxes determined.  The simulation 
exposures were set to 500\,ks to improve the statistics.  The fluxes of 
the residual signals were then determined from the simulated data.  The 
contributions were modeled as the best-fit absorbed power law 
model (but to the residual flux) in the spectra analysis.

The CXB was modeled as a power law with a fixed photon index of 1.4 
\citep{KIM+02} and a thawed normalization.  Thawing the photon index 
gives a value of 1.39 (1.37) for the (de-)projection spectral analyisis,
and this systematic uncertainty was taken into account in our results 
(Appendix~\ref{sec:app1}).
We determined that the surface brightness of this residual CXB is 
generally $S_{\rm CXB,{\it Suzaku}}$\,=\,(1.2--$1.8)\times 
10^{-11}$\,erg\,cm$^{-2}$\,s$^{-1}$\,deg$^{-2}$ in 2--10\,keV in the outer 
regions beyond 27\arcmin.  Since our point source detection limit is 
about $1.4 \times 10^{-14}$\,erg\,cm$^{-2}$\,s$^{-1}$ in this	 
band, the expected level of the unresolved CXB is estimated to be 
$S_{\rm CXB,expect}$\,$\approx$\,$1.3\times 10^{-11}$\,erg\,cm$^{-2}$\,s$^{-1}$\,deg$^{-2}$ with a 
$1\sigma$
cosmic variance of $\sigma_S$\,$\approx$\,$1.1\times 10^{-12} (\Omega/150\,{\rm 
arcmin}^{2})^{-0.5}$\,erg\,cm$^{-2}$\,s$^{-1}$\,deg$^{-2}$ 
\citep{MCL+03,Mor+09}, where $\Omega$ is the solid angle of the spectral 
extraction regions, which is typically between 100 and 200~arcmin$^{2}$ 
beyond 27\arcmin.  Thus, the measured residual CXB is in a very good 
agreement with the expected value.

We modeled the GXB as an absorbed two-thermal-component \citep[{\tt 
APEC};][]{SBL+01}: one for the cool halo and the other for the hot halo.  
We fixed the metallicity at one solar, the redshift at zero, and the 
absorption to the Galactic value \citep{WSB+13}.  The two temperatures 
were each thawed and tied in all regions.  The best-fit temperatures are 
$T_{\rm CH}$\,$\approx$\,$0.14$\,keV and $T_{\rm HH}$\,$\approx$\,$0.5$\,keV for the 
cool halo and hot halo, respectively.  The Antlia Cluster is located at 
a low Galactic latitude of 19\arcdeg; such a high $T_{\rm HH}$ has been 
observed in other low Galactic latitude observations \citep{Yos+09}.  We 
have checked the systematic uncertainty of the hot halo temperature by 
fixing $T_{\rm HH}$ at 0.3\,keV, and this uncertainty was included 
in our analysis (Appendix~\ref{sec:app1}). The normalizations of the 
cool and hot halos were thawed and untied in all regions to take into 
account the angular variation.  We do not include the very soft 
$\sim$0.08\,keV component for the Local Bubble, because it was not detected in our 
data analysis, possibly due to the 0.6\,keV cut-off energy we used.
\citet{SVD+09} found a correlation between the GXB surface brightness 
and the {\it ROSAT} RASS R45 flux (in {\it ROSAT} PI channels 52--90).  
The R45 
flux in an annulus of 1\arcdeg--2\arcdeg\ centered at Antlia is $211 
\times 10^{-6}$ counts\,s$^{-1}$\,arcmin$^{-2}$, while in a circular 
region of a 0\fdg2 radius centered at the EB field is $254 \times 
10^{-6}$ counts\,s$^{-1}$\,arcmin$^{-2}$.  From spectral analysis, we 
measured the GXB surface brightness to be (1.1--$1.2) \times 
10^{-11}$\,erg\,s$^{-1}$\,cm$^{-2}$\,deg$^{-2}$ in 0.47--1.21\,keV beyond 
1\arcdeg\ from Antlia, which is in excellent agreement with the expected value shown 
in Figure~2 of \citet{SVD+09}.

SWCX mainly produces line emission in the soft band $\lesssim 2$\,keV. We 
have filtered out soft X-ray flares potentially caused by SWCX, but the 
steady emission might also be contaminated by a constant level of SWCX.  
Thus, we include eight Gaussian models to take into account for the SWCX 
emission lines between 0.6--2.0\,keV.  The energies of the lines were 
fixed to those in Table~2 of \citet{Bau+09}.  The Gaussian widths were fixed 
to zero.  All the normalizations were thawed.

\begin{figure}
\includegraphics[width=0.4\textwidth,angle=270]{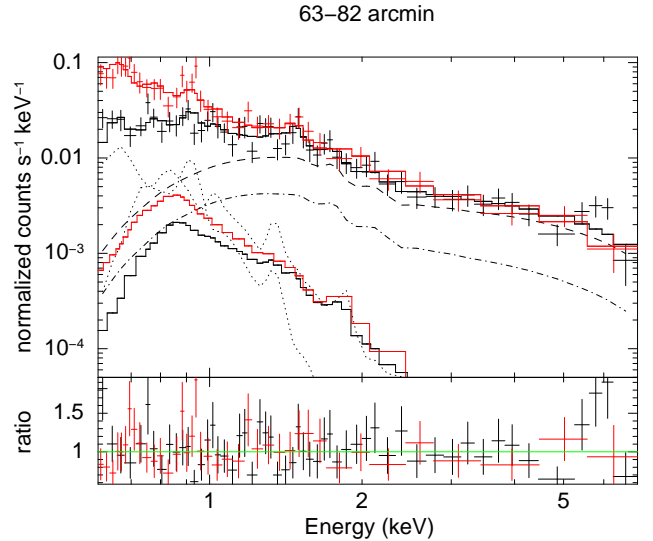}
\caption{
Representive {\it Suzaku} spectra covering the 
region around $R_{200}$.
The FI and XIS1 data are shown in black and red, respectively.
The solid lines following the data are the sum of all the model components.
The two solid lines below the data indicate the ICM component.
For clarity, we only show the GXB (the two dotted lines), the CXB 
(dashed), and the residual signal from the removed point sources 
(dash-dotted) components for the FI data.
The error bars of the data are at 1$\sigma$.
}
\label{fig:spec}
\end{figure}

In the spectral analysis, we included both the ICM and background models 
for all regions, except for the local background field (EB) where only 
the background model was used.  We simultaneously fitted the spectra 
from all the different regions 
in all the pointings (E0--EB).  
The unresolved CXB and SWCX are subject to cosmic variance and time 
variation, respectively.
Therefore, we did not tie the CXB and SWCX normalizations for 
different regions, but only tied them for different detectors in the same 
spectral region.
The ICM temperature, metallicity, and normalization were also not tied 
for different regions, but tied for the different detectors in the the 
same spectral region, with the metallicity beyond 18\arcmin\ fixed to a 
constant value as mentioned above.
Representive spectra covering the region around $R_{200}$ are 
shown in Figure~\ref{fig:spec}.

\section{Surface Brightness Profile}
\label{sec:SB}

\begin{figure}
\includegraphics[width=0.4\textwidth,angle=270]{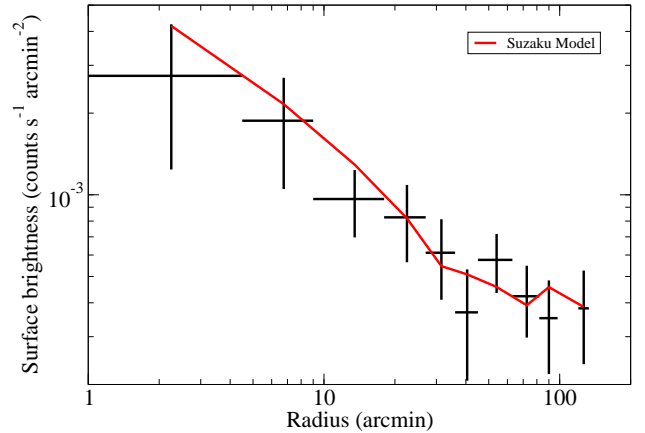}
\caption{
RASS surface brightness profile in the standard R47 band 
($\sim$0.5--2\,keV) extracted with the same {\it Suzaku} regions is shown in 
black data points.  The red line is the simulated RASS profile using the 
best-fit {\it Suzaku} model and the ROSAT spectral response.
}
\label{fig:RASS}
\end{figure}

Soft X-ray emission can be strongly enhanced by SWCX or particle-induced 
background.  Calibration uncertainty of the response files can also bias 
the emission measurement.  To test whether there is significant 
systematic uncertainty in the soft X-ray emission measured with {\it 
Suzaku}, we have extracted the surface brightness profile from the ROSAT 
X-ray All-Sky Survey (RASS) using the same regions as the {\it Suzaku} 
analysis (Figure~\ref{fig:RASS}).  We simulated the RASS surface 
brightness profile using the ROSAT response file and the best-fit {\it 
Suzaku} model.  The simulated profile is consistent with the RASS data, 
suggesting that the soft X-ray emission measured with {\it Suzaku} is 
reliable.

\begin{figure*}
\includegraphics[width=0.4\textwidth,angle=270]{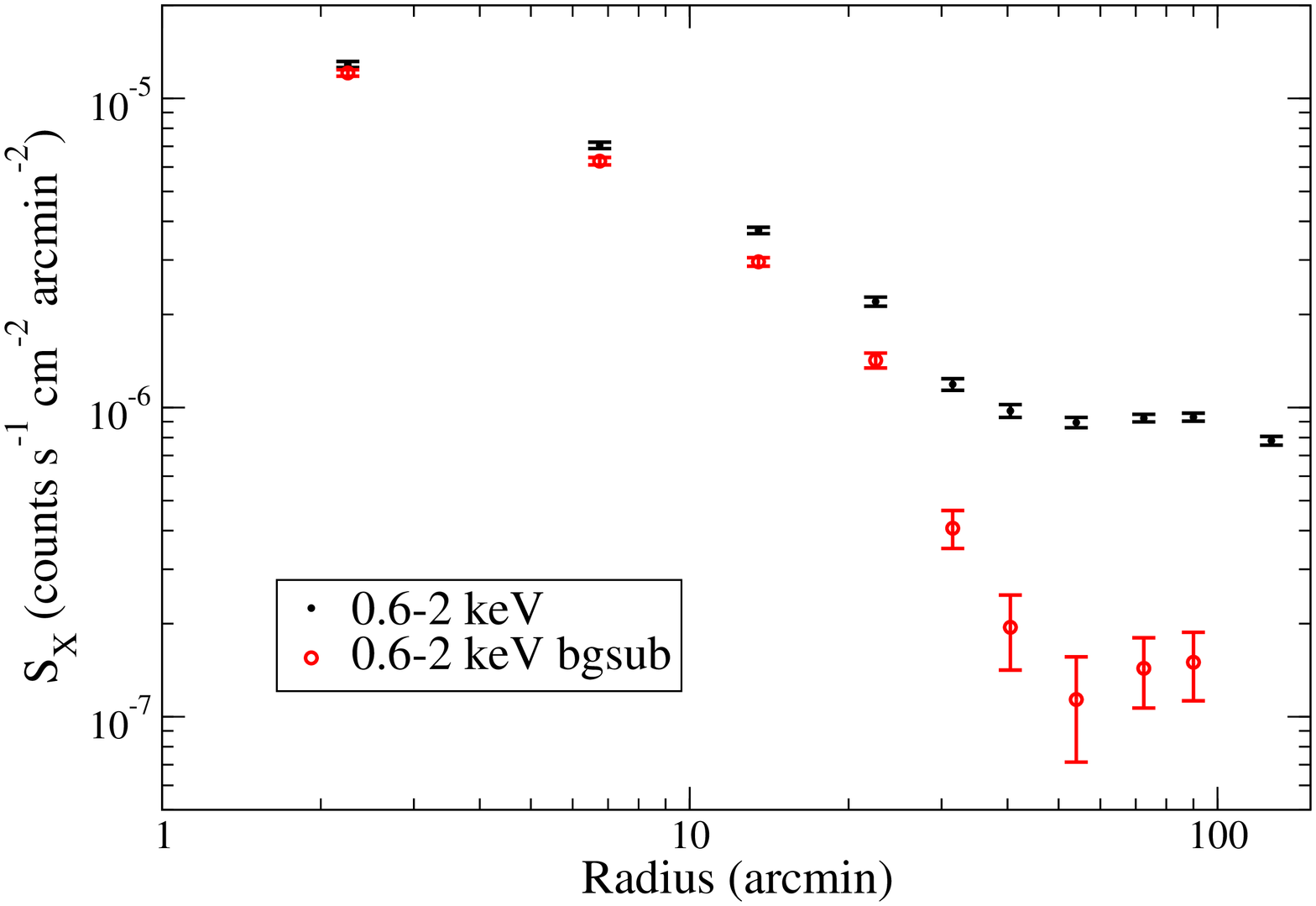}
\includegraphics[width=0.4\textwidth,angle=270]{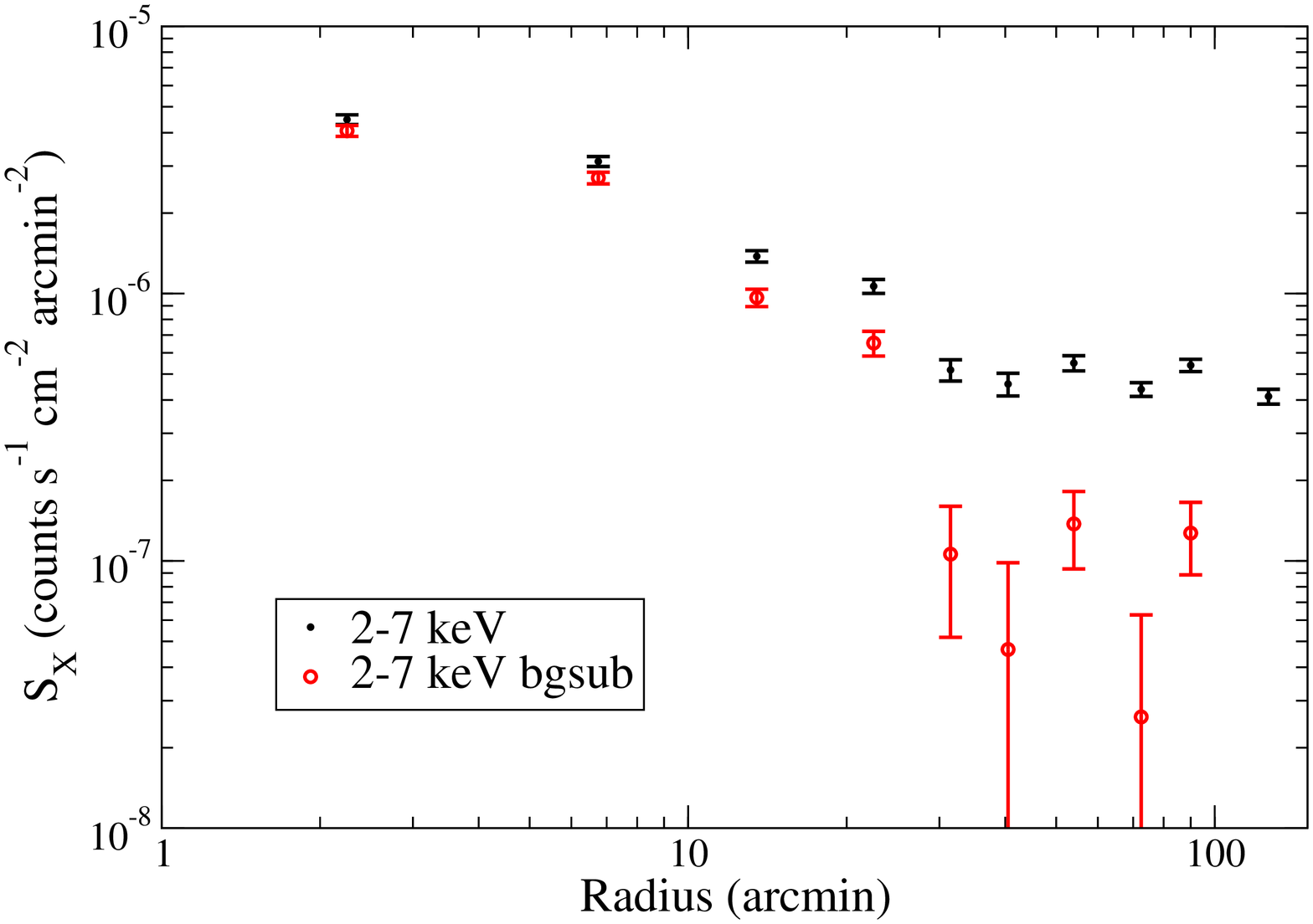}
\caption{
Left panel: surface brightness profiles of the unresolved diffuse 
emission in the 0.6--2\,keV band (black solid circles).  
Red open circles represent
the profile with local background subtracted.
Errors of the local background have been added in quadrature.
Right panel: similar to the left panel but in the 2--7\,keV band.  
}
\label{fig:surbri}
\end{figure*}

The surface brightness profiles in the soft (0.6--2\,keV) and hard 
(2--7\,keV) energy bands are shown as black solid circles in 
Figure~\ref{fig:surbri}.  NXB has been subtracted in these profiles.  We 
also subtracted the local background (EB: the last data bin), and created 
the corresponding local background subtracted surface brightness 
profiles (red open circles), while the error of the local background has 
been added in quadrature.  The local background subtracted surface 
brightness profile of the soft band (0.6--2\,keV) decreases from the 
center to a minimum value at about 54\arcmin, and then flattens beyond 
that.  The soft emission at the outermost two bins 
(63\arcmin--98\arcmin) is each above the local background by about 
7$\sigma$.  Spectral analysis suggests 
that the flattening emission is at least partially caused by the 
slight enhancement of SWCX, 
and therefore the ICM contribution in the 
outer regions cannot be determined from this surface brightness profile. 
The local background subtracted surface brightness profile of the hard 
band (2--7\,keV) fluctuates around a small value beyond 
$\sim$20\arcmin--30\arcmin, where the CXB dominates the emission.

\begin{figure*}
\includegraphics[width=0.4\textwidth,angle=270]{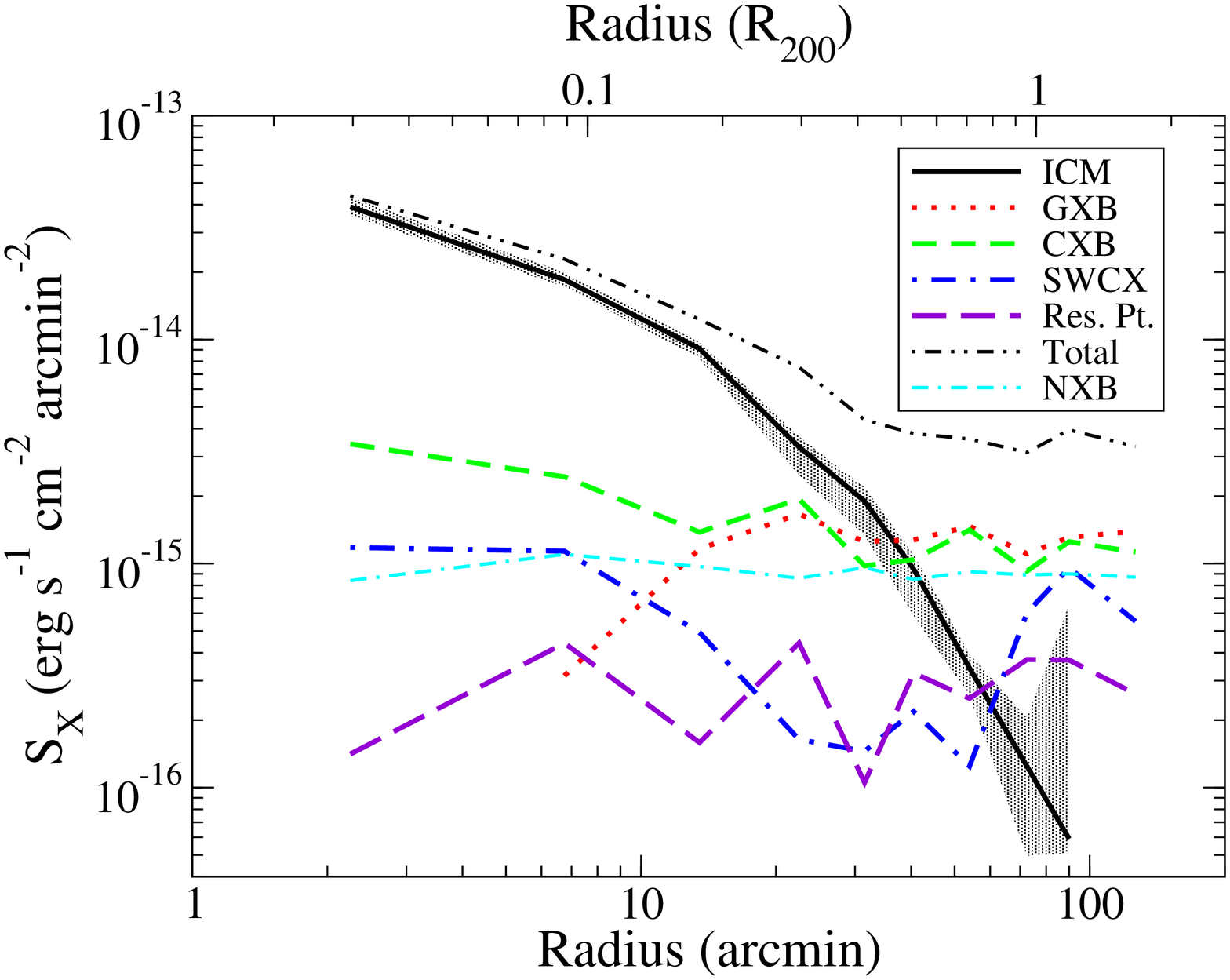}
\includegraphics[width=0.4\textwidth,angle=270]{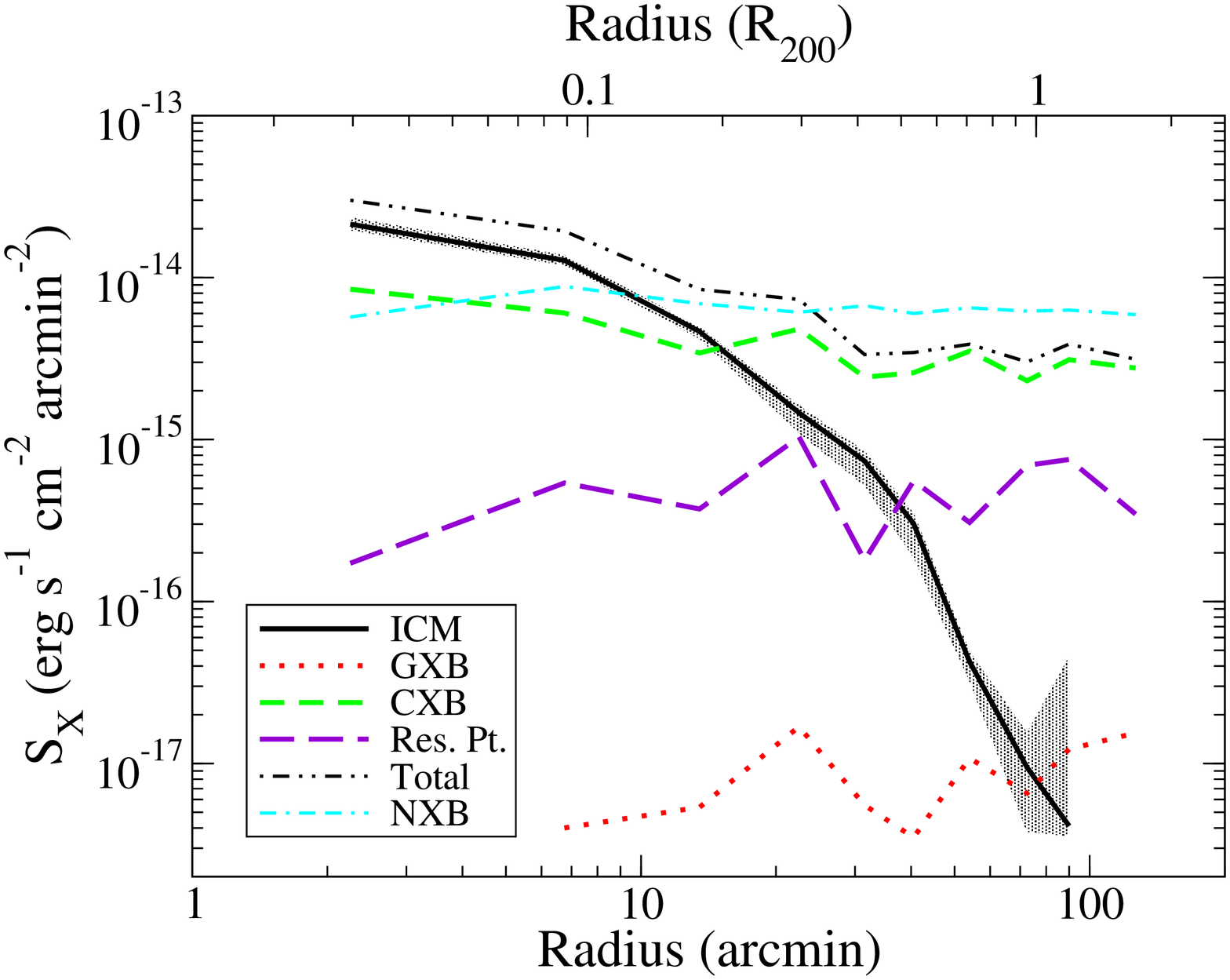}
\caption{
Left panel: surface brightness profiles of different components
from 0.6--2\,keV determined from spectral fitting.  
For the ICM component, the solid line is the best-fit profile and the 
shaded regions correspond to its confidence regions, including 
both statistical and systematic errors.
The ``Total'' profile is the sum of all components except the 
NXB.
For clarity, only the NXB of the XIS1 detector, which has a higher noise 
level than the XIS0 or XIS3 detectors, is plotted.
Right panel: same as the left panel but from 2--7\,keV.  
}
\label{fig:surbri_fit}
\end{figure*}

In the left panel of Figure~\ref{fig:surbri_fit}, 
the surface brightness profiles of different components in the soft band 
(0.6--2\,keV)
calculated using the spectral fitting models are shown.  
We also show the NXB component, which was subtracted from the 
data before the spectral fitting.  For clarity, only the
NXB of the XIS1 detector, which has a higher noise level than 
the XIS0 or XIS3 detectors, is plotted.
For the ICM 
component, the solid line is the best-fit profile and the shaded regions 
correspond to its confidence regions calculated using the uncertainties 
of the {\tt APEC} normalizations (Appendix~\ref{sec:EM}); 
both statistical and systematic errors 
are included.  
Both the GXB and CXB dominate the soft emission in the outer regions
and are flat in the outer regions, 
while the residual signal from removed point sources (Res. Pt. in 
Figure~\ref{fig:surbri_fit}) is typically the least dominant component 
everywhere.
The NXB is almost always lower than the GXB and CXB, and its 
systematic uncertainty is $\lesssim 5\%$.
At radii beyond $\sim$54\arcmin, the surface brightness 
of the SWCX is comparable to the ICM.  The SWCX is in fact enhanced beyond 
that radius.  
This is supported by the fact that the 
solar proton fluxes measured by {\it ACE} are about 
a factor of two to eight higher during the observations for regions
beyond 63\arcmin\ compared to those between 
9\arcmin\ and 45\arcmin.\footnote{No proton flux
data were available for the observations $<$9\arcmin\ and between 
45\arcmin\ and 63\arcmin.}
Ignoring the SWCX component will overestimate the ICM 
emission in the outer regions.  

The right panel of 
Figure~\ref{fig:surbri_fit} shows that the hard-band (2--7\,keV) surface 
brightness profiles.
The NXB is the dominant component.
The NXB subtracted profile (``Total'' in the figure) 
is almost completely dominated by CXB beyond 
$\gtrsim$20\arcmin--30\arcmin.  Nevertheless, the very hard and unresolved CXB 
can be separated from the rest of the soft emission spectroscopically.

\section{Temperature Profile}
\label{sec:tprofile}

\begin{figure}
\includegraphics[width=0.4\textwidth,angle=270]{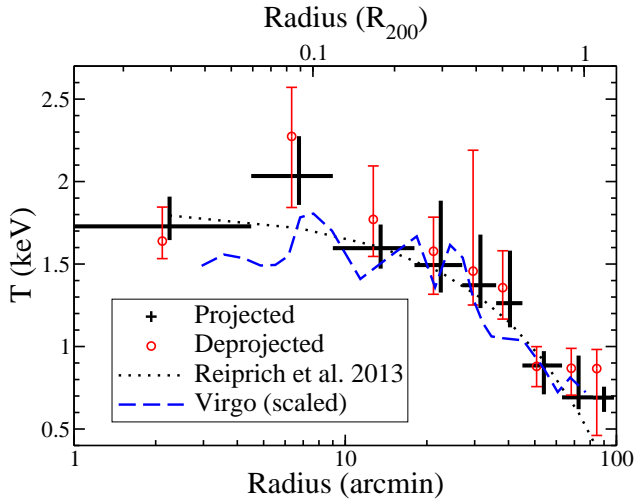}
\caption{
Projected (black crosses) and deprojected (red circles) 
temperature profiles of Antlia.  
Error bars for temperature include 
both statistical and systematic uncertainties.
The deprojected temperature data points have been slightly shifted to the
left for clarity.
The dotted line is the average scaled profile of a sample of clusters 
measured with {\it Suzaku} \citep{RBE+13}. The blue dashed line is the 
Virgo profile with its projected temperature scaled by a factor of $\langle T 
({\rm Antlia}) \rangle / \langle T ({\rm Virgo}) \rangle = 1.54/2.3$ and 
radius in units of $R_{200}({\rm Virgo}) = 234\arcmin$ \citep{UWS+11}.
Note that the actual temperature of 
Virgo drops from its peak by a factor of three near the center ($\lesssim 
10^{-3} R_{200}$), which cannot be seen on the radial scale of this 
figure. 
}
\label{fig:t_profile}
\end{figure}

The projected and deprojected temperature profiles are shown in 
Figure~\ref{fig:t_profile}, with temperature error bars including both 
statistical and systematic uncertainties (Appendix~\ref{sec:app1}).  The 
projected temperature drops from the central region of $\sim$2 to 
$\sim$0.7\,keV at the edge of the detection region of $\sim$100\arcmin\ 
(1.2~Mpc).  Such a declining temperature is typical in other clusters 
and groups.  The nearly isothermal central region, the slowly decreasing 
outer temperature, and the smaller emission contribution from the outer 
regions ensure that the projected temperature profile is similar
to the true deprojected temperature profile.  

We fitted the global temperature between $\sim$0.2--1~$R_{500}$ for
pie regions with the same angular width, and the 
best-fit temperature is $1.54^{+0.27}_{-0.13}$\,keV.  This global 
temperature is in fact used to determine the radius scale used 
throughout the paper (Section~\ref{sec:intro}).

The temperature profiles of most of the regular or less disturbed 
clusters appear to be self-similar beyond $\gtrsim 0.3 R_{200}$ 
\citep[e.g.,][]{RBE+13}.  In Figure~\ref{fig:t_profile}, we overplot 
the average scaled projected temperature profile measured with {\it 
Suzaku} and compiled by \citet{RBE+13} using a sample of 18 clusters 
with the following form:
\begin{equation}
\label{eq:scaledT}
k_{\rm B} T / \langle k_{\rm B} T \rangle = 1.19 - 0.84 R/R_{200},
\end{equation}
where we used the average Antlia temperature $\langle k_{\rm B} T 
\rangle = 1.54$\,keV and $R_{200}$\,=\,76\arcmin.  The projected 
temperature profile 
of Antlia agrees very well with the average scaled temperature profile 
from the center out to $\sim$$R_{200}$.  The projected temperature at about 
$1.2$$R_{200} \approx 90\arcmin$ is a factor of 2.3 higher than the average 
scaled profile, which can be explained by the 
local deviation in the temperature
or the large scatter of temperature seen in the sample of \citet{RBE+13}.
For comparison, we also plot the 
projected temperature profile of Virgo measured by \citet{UWS+11} using {\it 
XMM-Newton}, with its temperature scaled to match those 
of the Antlia Cluster.  Other than the small-scale fluctuations caused 
by substructures, the general shape of the Virgo temperature profile 
agrees very well with that of the Antlia out to $\sim$$R_{200}$, 
as well as the average scaled 
profile.  The agreement of these two low temperature groups with the 
sample of \citet{RBE+13}, which includes mostly massive clusters 
$>$4\,keV, suggests that the temperature profile of low mass groups may 
also follow the self-similar profile of the more massive clusters beyond 
the core regions out to $R_{200}$.  
It should be noted that both Antlia 
and Virgo are measured in only one direction.  More observations of galaxy 
groups will be needed to test the self-similarity near $R_{200}$.

\section{Density, Pressure, and Entropy Profiles}
\label{sec:DPE}

\begin{figure}
\includegraphics[width=3.5in, angle=0]{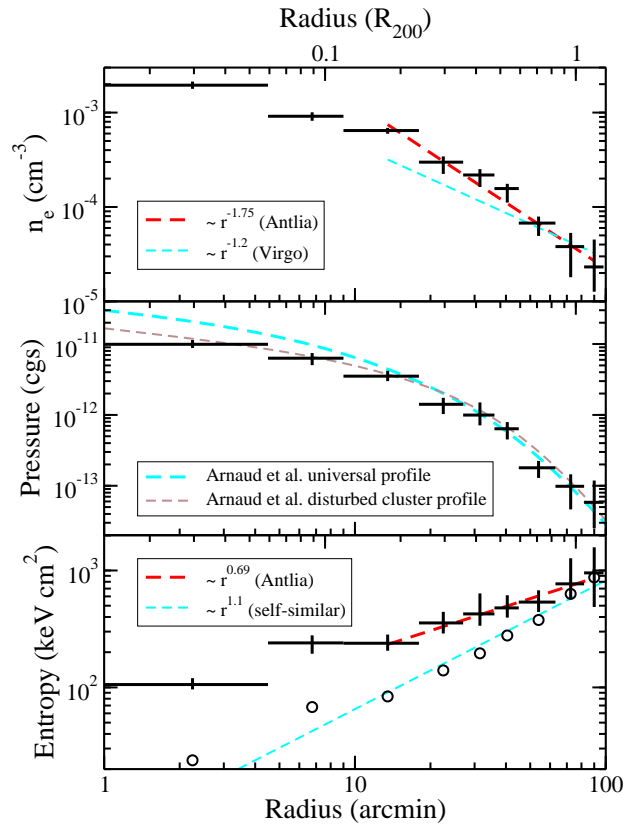}
\caption{
Upper panel: electron density profile of Antlia
(black data), with a power law index of 
$\alpha = 1.75^{+0.27}_{-0.24}$ beyond 
$\sim$10\arcmin\ (thick red dashed).  The thin cyan dashed line is the best-fit
power law for Virgo ($\alpha = 1.2\pm0.2$) plotted in units of its $R_{200}$ 
\citep{UWS+11}.
Middle panel: gas pressure profile of Antlia (black data).  The standard 
universal
pressure profile of clusters (thick cyan dashed) and 
the version with only morphologically disturbed clusters (thin brown dashed) 
are shown.
Lower panel: entropy profile of Antlia (black crosses), with a power law 
index of
$0.69^{+0.22}_{-0.24}$ beyond the core of 
$\sim$10\arcmin\ (thick red dashed).  The thin cyan dashed line is the 
gravity heating-only model with a power law index of 1.1.
The circles are measured entropy multiplied by the gas correction factor 
of $[f_{\rm gas}(r)/0.15)]^{2/3}$ (see text).  
Errors bars in density, pressure, and entropy include both statistical 
and systematic uncertainties.
}
\label{fig:npeprofile}
\end{figure}

We used both the projected and deprojected spectral normalizations of 
the ICM component to constrain the electron density (Appendix~\ref{sec:EM}).
The resulting electron density profile is shown in the upper panel of 
Figure~\ref{fig:npeprofile}.

The density profile has a rather flat core within 
$\sim$10\arcmin--20\arcmin\ and steepens beyond that.  
The density decreases all the way beyond $R_{200}$ out to $\sim$100\arcmin.
We 
fitted a power law to the density profile ($n_e \propto r^{-\alpha}$) 
beyond $\sim$10\arcmin, and the power law index is $\alpha = 
1.75^{+0.27}_{-0.24}$.  This is consistent with the density slope of 
1.65--2.25 at $R_{500}$ measured with 43 nearby galaxy groups using {\it 
Chandra} \citep{SVD+09}.
It is not as steep as those of more 
massive galaxy clusters, which have a slope of $\sim$2 at $R_{200}$ or 
even 3 at radii $\gtrsim R_{200}$ \citep[e.g.,][]{MSF+15}. The density 
profile of Antlia in its outer regions is significantly steeper than that of 
Virgo \citep[$\alpha_{\rm Virgo} = 1.21\pm0.20$\footnote{The error of the Virgo 
density index has been converted to 90\% confidence.};][]{UWS+11}.

With the deprojected temperature and density profiles measured, we 
calculated the gas pressure profile, $P = nkT$, where $n$ is the total 
number density in the gas (both ions and electrons).  We assume $n 
\approx 1.92 n_e$ for a fully ionized ICM.  The pressure profile (middle 
panel in Figure~\ref{fig:npeprofile}) in the central $\sim$30\arcmin\ is 
similar to the density profile because the temperature is quite uniform.  
Beyond that, the pressure profile is steeper due to the declining 
temperature.

\citet{APP+10} found that more massive galaxy clusters with $M_{500} 
\gtrsim 10^{14}M_{\odot}$ (or $T_X \gtrsim 2$\,keV) obey the universal 
pressure profile of the form:
\begin{equation}
\label{eq:univpress}
\frac{P(x)}{P_{500,\rm ad}} = \frac{P_0}{(c_{500}x)^{\gamma}[1+(c_{500}x)^{\alpha}]^{(\beta-\gamma)/\alpha}},
\end{equation}
with the characteristic pressure adjusted for the slight deviation from self-similar given by
\begin{multline}
\label{eq:P500ad}
P_{500,\rm ad}=1.65\times10^{-3} E(z)^{8/3}
\\
\times \left[
\frac{M_{500}}{3\times10^{14}h^{-1}_{70} M_{\odot}}\right]^{2/3+\alpha(x) } h^2_{70}$~keV~cm$^{-3},
\end{multline}
where $\alpha(x) = 0.22/[1+(2x)^3]$ accounts for the breakdown of 
self-similarity, $x\equiv r/R_{500}$, and $E(z) = \sqrt{\Omega_M 
(1+z)^3+\Omega_{\Lambda}}$.  \citet{SSV+11} found that the universal 
pressure profile can be extended to less massive groups. The thick cyan 
dashed line in the middle panel of Figure~\ref{fig:npeprofile} shows the 
universal profile with the best-fitted parameters for all the cluster 
sample of 
\citet[][i.e., [$P_0, c_{500}, \gamma,\alpha,
\beta${]}\,= [$8.403h_{70}^{-3/2}, 
1.177, 0.3081, 1.0510, 5.4905${]}]{APP+10},
with the radius scale of the Antlia Cluster $R_{500}$\,=\,591\,kpc.  
The pressure profile of Antlia agrees very well with the 
universal profile beyond $\sim$30\arcmin, with a central deficit in 
pressure within that region.  If we use the parameters for the 
morphologically disturbed clusters
\citep[i.e., $[P_0, c_{500},
\gamma,\alpha,\beta{]}=[3.202h_{70}^{-3/2}, 1.083, 0.3798, 1.4063, 5.49{]}$;
thin brown dashed line;][]{APP+10},
the pressure agrees over the full 
radial range.  
\citet{Urb+14} showed that in Perseus, where the directions are 
strongly disturbed by 
cold fronts or sloshing, the pressure profiles deviate significantly from 
the universal profile.
The agreement in Antlia (in particular beyond $\sim$30\arcmin)
suggests that the gas profile along the 
east direction may be a fair representation of the azimuthally averaged 
profile.

We also calculated the gas entropy parameter $K \equiv k_{\rm B} 
T/n_e^{2/3}$, which reflects the thermodynamic history of the hot plasma 
(bottom panel in Figure~\ref{fig:npeprofile}). A self-similar model with 
only gravitational collapse heating predicts the entropy profile to be 
\citep{VKB05,Pra+10}:
\begin{equation}
\label{eq:entropy}
K = 1.32 K_{200,\rm adi} \left(\frac{r}{r_{200}}\right)^{1.1},
\end{equation}
where 
\begin{equation}
\label{eq:K200}
K_{200,\rm adi} = 362 {\rm \,keV\,cm}^2 \left(\frac{\bar{T}_{200}}{1 {\rm 
keV}}\right)
\left(\frac{0.15}{f_{\rm b}}\right)^{2/3} E(z)^{-4/3},
\end{equation}
where $f_{\rm b}$ is the cosmic baryon fraction.
The characteristic temperature here is related to the virial mass by 
$k_{\rm B} \bar{T}_{200} \equiv G M_{200} \mu m_p / 2R_{200}$,
where $\mu$ is the mean molecular weight per particle and $m_p$ is the
proton mass.  This 
self-similar model with the radius scale of the Antlia Cluster $R_{200} 
= 887$\,kpc is plotted as a thin cyan dashed line.  The central entropy 
of cool core clusters can be biased low, while it can be biased high for 
non-cool core clusters.  We therefore fitted a power law to the Antlia 
entropy profile excluding the $\sim$10\arcmin\ core.  The best-fit 
power law index is $0.69^{+0.22}_{-0.24}$, which is significantly flatter than 
that of $K \propto r^{1.1}$ for the self-similar model with only 
gravitational collapse heating.  
The entropy appears to be increasing from the center all the 
way out to $\sim$$R_{200}$.
Going further, 
there is no evidence of entropy flattening or dropping beyond $R_{200}$,
although the uncertainty is too large to be conclusive.

Within the $\sim$10\arcmin\ core, the entropy is significantly higher than
the self-similar model, which is typical for a non-cool core cluster.
The overall profile is also clearly flatter than 
the self-similar model.
Note that 
the magnitude of entropy beyond $R_{500} = 51\arcmin$ is in fact 
consistent with the gravity heating-only model.  
The flattening in 
Antlia should thus be caused by the increase of entropy inside that 
radius, which is different from other more massive clusters, where the 
flattening is caused by a lower entropy beyond $\sim$$R_{500}$ 
\citep[e.g.,][]{WFS+13}.  
The entropy of Antlia 
at $R_{500(2500)}$\footnote{$R_{2500} \sim 0.5 R_{500}$ for Antlia, as well 
as many nearby galaxy groups \citep{SVD+09}.} is about 50(100)\% 
higher than the gravity-only self-similar model, which is similar to typical 
galaxy groups found by \citet{SVD+09}.  
The extra entropy may be caused by any previous AGN feedback in 
the past, supernova feedback, preheating of gas before accretion, 
conduction that transfers heat from outer regions, low 
entropy gas cooling out of the hot phase, or some combination of
these processes \citep[e.g.,][]{SVD+09,Pra+10}.
At large radii beyond 
$\sim$$R_{500}$, there is no evidence of the entropy changing due to 
non-gravitational processes for Antlia.
Note that due to the large temperature uncertainty, 
if the temperature beyond $R_{200}$ is biased high by a 
factor of two, the entropy will be reduced by the same factor, and the 
entropy profile can be flat beyond that.

\citet{Pra+10} found that the general existence of a central entropy 
excess in clusters is connected to the gas faction.  By introducing a 
gas correction factor, the corrected entropy matches the 
theoretical prediction better.  In the lower panel of 
Figure~\ref{fig:npeprofile}, the circles represent the measured entropy 
multiplied by a factor of $[f_{\rm gas}(r)/0.15]^{2/3}$, where $f_{\rm 
gas}(r)$ is the gas fraction measured in Section~\ref{sec:mass} below.  
The corrected entropy profile of Antlia becomes largely consistent with 
the theoretical prediction.

\section{Gas and Hydrostatic Mass Profiles}
\label{sec:mass}

\begin{figure}
\includegraphics[width=0.5\textwidth]{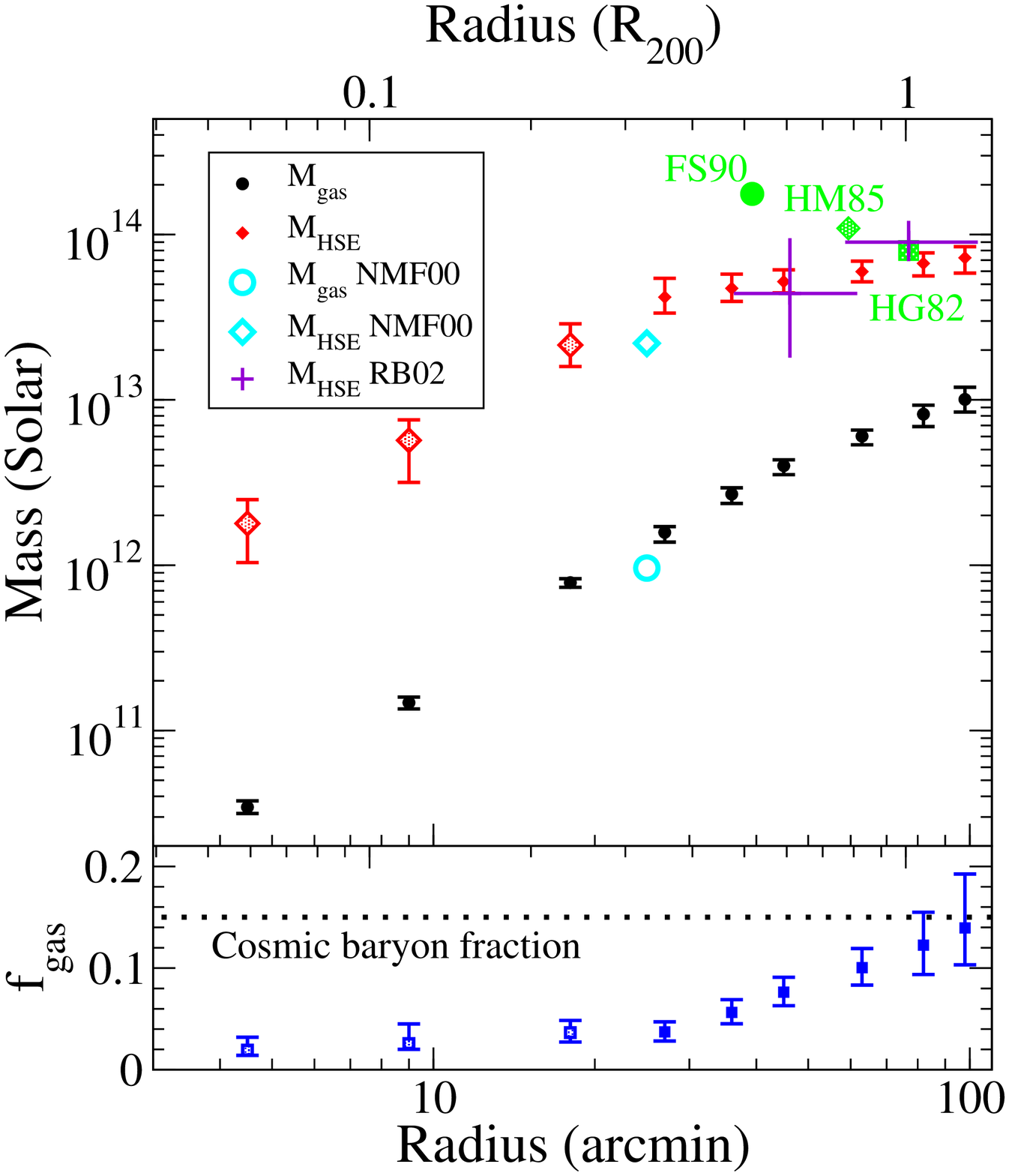}
\caption{
Upper panel: gas-mass $M_{\rm gas}$ (small solid black circle) and HSE 
mass $M_{\rm HSE}$ (large red diamond using equation~(\ref{eq:mhse});
small solid red diamond using equation~(\ref{eq:mhse2})) 
profiles of Antlia.  X-ray 
measurements by \citet[][NMF00]{NMF+00} are shown with large open cyan circles 
and 
diamonds.  $M_{\rm HSE}$ estimated by \citet[][RB02]{RB02} is shown with violet 
crosses, with error bars scaled to 90\% confidence.  Dynamical mass 
estimations measured by \citet[][HG82]{HG82}, \citet[][HM85]{HM85}, 
and \citet[][FS90]{FS90} are shown as shaded green points. All of the 
measurements have been rescaled to the Hubble constant or distance we 
adopted. Lower panel: gas fraction [$f_{\rm gas}(<r)\equiv M_{\rm 
gas}/M_{\rm HSE}$] of Antlia.  The horizontal dashed line shows the 
comic baryon fraction of 0.15.
Note that, as usual throughout the paper, 
the scaled radius $R_{200}$ on the top-axis 
is defined using the scaling relation (Section~\ref{sec:intro}) 
instead of using the 
HSE mass determined in this section.
}
\label{fig:mass}
\end{figure}

With the density profile measured, we can calculate the 
enclosed gas-mass profile by
\begin{equation}
M_{\rm gas}(< r)
=
4 \pi \int_0^r d r^\prime r^{\prime 2} \rho_{\rm gas} ( r^\prime )
\, ,
\label{eq:mgas}
\end{equation}
where
$\rho_{\rm gas} = \mu_e m_p n_e$ is the gas mass density, and $\mu_e$ is 
the mean molecular weight per electron determined from the gas 
abundances.  We also calculated the HSE mass given by 
\citep[e.g.,][]{Sar86}
\begin{equation}
M_{\rm HSE}(< r)
=
-\frac{kTr}{\mu m_p G}
\left( \frac{d \ln \rho_{\rm gas}}{d \ln r} + \frac{d \ln T}{d \ln r}  \right)
\, .
\label{eq:mhse}
\end{equation}
For regions inside $\sim$20\arcmin, we directly applied 
equation~(\ref{eq:mhse}) on the measured density and temperature to 
calculate $M_{\rm HSE}$.  However, for regions beyond that, the 
uncertainty is very large due to the large uncertainties in the data.
To capture the global behavior, we fitted 
power law models to the temperature and density profiles beyond 
$\sim$20\arcmin, $T \propto r^{\Gamma_T}$ and $\rho \propto 
r^{\Gamma_\rho}$; therefore equation~(\ref{eq:mhse}) becomes
\begin{equation}
M_{\rm HSE}(< r)
=
-\frac{kTr}{\mu m_p G}
\left( \Gamma_T + \Gamma_\rho \right)
\, .
\label{eq:mhse2}
\end{equation}
In calculating $M_{\rm gas}$ and $M_{\rm HSE}$, we used the $10^6$ 
simulated density and temperature profiles, and assessed the median and 
errors in a Monte Carlo sense described in Appendix~\ref{sec:EM}.  
The mass profiles are shown in Figure~\ref{fig:mass}.

The gas-mass increases from $3.4\times 10^{10}M_{\odot}$ at 4\farcm5 
to $4.6~(7.5)\times 10^{12}M_{\odot}$ at $R_{500(200)} = 51\arcmin\ 
(76\arcmin)$.  Compared with the gas mass, the HSE mass increases with a 
flatter slope, with $M_{\rm HSE}(R_{500})=5.5\times 10^{13}M_{\odot}$ 
and $M_{\rm HSE}(R_{200})=6.5\times 10^{13}M_{\odot}$.  If we use our 
adopted scale radius to calculate the scale mass $M_{\Delta} \equiv 4\pi 
\Delta \rho_c R_{\Delta}^3/3$, we find that $M_{500(200)} = 5.9~(7.9) 
\times 10^{13}M_{\odot}$, which is 7 (20)\% higher than $M_{\rm HSE}$ 
at $R_{500(200)}$.  While the precise HSE mass bias from the total mass 
is uncertain, it can be biased low by 10--40\% beyond $R_{500}$ 
\citep[e.g.,][]{MHB+13,Oka+14}.  
Because we only have measurements in one direction,
azimuthal variation may 
also bias these mass estimations on the 
same order.  
Since we found that the HSE masses $M_{\rm HSE}$ are 
 7\% and 20\% 
lower than
the scaled masses $M_{\Delta}$ calculated using our adopted 
scaled radii at $R_{500}$ and $R_{200}$, respectively, 
the deviations are insignificant compared to the mass biases 
introduced by azimuthal variation and HSE bias.\footnote{Note that 
the scaled masses $M_{\Delta}$ are adopted using the scaling 
relation of \citet{SVD+09}, 
while this scaling relation is also measured using the HSE method 
and is thus subjected to HSE bias.}
Thus, we can adopt the
scaled radii using the X-ray 
scaling relation.

In Figure~\ref{fig:mass}, we also plot other mass measurements in the 
literature.  All of the measurements have been rescaled with the Hubble 
constant or distance we used.  Compared to previous X-ray measurements 
with ROSAT and ASCA \citep{NMF+00} at $\sim$$0.5R_{500}$, our $M_{\rm 
gas}$ and $M_{\rm HSE}$ are about 40\% and 60\% higher, respectively.  
The difference 
in $M_{\rm gas}$ might be partially due to their high assumed metallicity of 
$0.35\, Z_{\odot}$, while our measured metallicity decreases from 
$\sim$$0.4\, Z_{\odot}$ down to $\lesssim 0.15\, Z_{\odot}$ at 
$\sim$$0.5R_{500}$.  
A bias high in metallicity results in a lower normalization in the 
spectral fitting, and hence a bias low in gas density or gas mass 
(Appendix~\ref{sec:app1}).
Another possible reason is due to their isothermal 
assumption when determining both $M_{\rm gas}$ and $M_{\rm HSE}$, as 
well as the prior assumption of a dark matter potential in their work.  
\citet{RB02} estimated the $M_{\rm HSE}$ at $R_{500}$, which is very 
consistent with our measurement.  Their $M_{\rm HSE}$ at $R_{200}$ is 
about 40\% higher.  This could be due to the isothermal temperature of 
1.18\,keV they assumed, which is biased high by a factor of about 1.2--2 between 
$R_{500}$ and $R_{200}$, and hence $M_{\rm HSE}$ could be biased high.  Mass 
estimations by assuming the optical galaxies in dynamical equilibrium at 
different characteristic radii are generally larger than our measured 
$M_{\rm HSE}$ \citep{HG82, HM85, FS90}.  However, the numbers of 
galaxies used in these measurements are very small ($\leq 21$), and 
therefore the statistical uncertainties are quite large.

The stellar mass can be estimated from the total K-band luminosity.  
\citet{LMS04} found that the K-band luminosity of all the galaxies 
inside $R_{200}$ is $L_{K,200} = 1.54\times10^{12} h_{70}^{-2} 
L_{\odot}$.\footnote{We rescaled the luminosity according to the 
distance we used.}  With a typical mass-to-light ratio of $M/L_K = 
0.95$ \citep{BMK+03}, the total stellar mass of the Antlia Cluster is 
therefore $M_{\star} = 1.5 \times10^{12} h_{70}^{-2} M_{\odot}$.  Thus, 
the gas starts to dominate over the stellar component for radii 
$\gtrsim$26\arcmin\,$\approx$\,0.34$R_{200}$.  
The stellar mass is about 30(20)\% of 
the gas mass at $R_{500(200)}$.

The gas-mass fraction is defined as $f_{\rm gas} \equiv M_{\rm gas} / 
M_{\rm HSE}$ and is shown in the lower panel of Figure~\ref{fig:mass}.  
The gas fraction increases from about 0.02 near the center to about 0.14 
at $\sim$100\arcmin.  The gas fraction does not exceed the cosmic baryon 
fraction\footnote{We adopted the cosmic baryon fraction of 
0.15 measured by 
{\it Planck} \citep{Planck}.} even slightly beyond $R_{200}$, suggesting 
that gas clumping or fluctuations are not significant in this direction of 
the Antlia Cluster.  Note that the HSE mass can be biased low by $>$30\% 
beyond $R_{500}$, and therefore the true gas fraction is likely to 
be even lower at the outer boundary \citep{Oka+14}.  The baryon 
fraction, which includes both the gas and stellar components, is at most 
a factor of 1.2 of the gas fraction at $R_{200}$, and thus will not 
alter our conclusion.

\section{Electron-ion equipartition and collisional ionization timescales}
\label{sec:time}

\begin{figure}
\includegraphics[width=0.4\textwidth, angle=270]{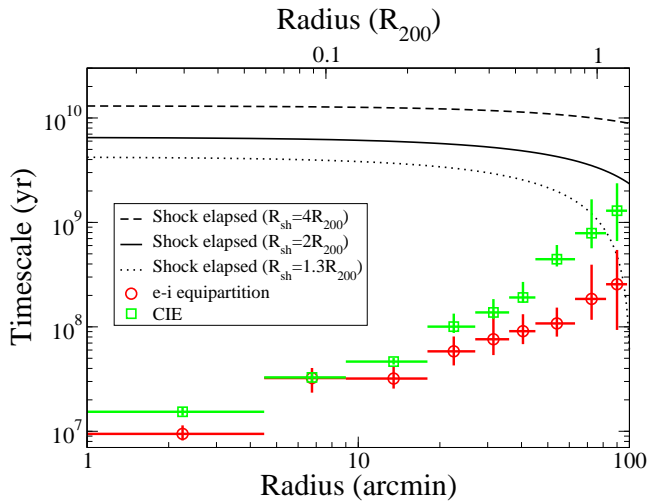}
\caption{
Electron-ion equipartition (red circles) and collisional ionization 
(green squares) timescales for Antlia.  The shock-elapsed timescales, 
assuming the shock radius is at 4$R_{200}$, 2$R_{200}$, and 1.3$R_{200}$ are 
shown in black dashed, solid, and dotted lines, respectively.
We expect ions and electrons to share the same temperature 
or ionization in CIE, where the equipartition or the collisional ionization
timescale is shorter than the shock-elapsed timescale, respectively.
}
\label{fig:time}
\end{figure}

Because of the low density in the outskirts of galaxy clusters/groups, the 
collisional timescales can be longer than the dynamical timescales.  
Thus, after hot gas has passed through an accretion shock, electrons and 
ions may not be in equipartition \citep[e.g.,][]{FL97,WS09} and 
collisional ionization equilibrium (CIE) may not be reached 
\citep{WSJ11}.  The electron-ion equipartition timescale is estimated 
to be \citep[][p.\,135]{Spi62}
\begin{equation}
\label{eq:tei}
t_{ei} \approx 7.0\times 10^8 {\rm \,year} \left(\frac{T_e}{10^7 {\rm \,K}} \right)^{3/2} \left(\frac{n_e}{10^{-5} {\rm \,cm}^{-3}} \right)^{-1}
\left(\frac{\ln \Lambda}{40} \right)^{-1},
\end{equation}
where $\ln \Lambda$ is the Coulomb logarithm.  The collisional 
ionization timescale for most elements of astrophysical interest is 
estimated to be \citep{SH10}
\begin{equation}
\label{eq:tcie}
t_{\rm CIE} \approx 3 \times 10^9 {\rm \,year} \left(\frac{n_e}{10^{-5} {\rm \,cm}^{-3}} \right)^{-1}.
\end{equation}
These two timescales of Antlia are plotted in Figure~\ref{fig:time}.  We 
compare these to the shock-elapsed timescale (i.e., the timescale since 
the gas has passed the accretion shock)
\begin{equation}
\label{eq:tsh}
t_{\rm sh}(r) = \frac{r - R_{\rm sh} }{v_{\rm sh}},
\end{equation}
where the shock velocity $v_{\rm sh} \approx v_{\rm infall}/3$.  The 
infalling velocity can be estimated as $ v_{\rm infall} \approx \sqrt{2 
k_{\rm B} T/\mu m_p}$ \citep[e.g.,][]{Tak98}.  We take the maximum 
temperature of Antlia to estimate the minimum $t_{\rm sh}$.  From 
numerical simulations, typical accretion shock radii are between $1.3 
R_{200} < R_{\rm sh} < 4 R_{200}$.  We adopted a shock radius of $ 
R_{\rm sh} = 2 R_{200}$ to be consistent with the simulations of
\citet{WS09}.  We also consider the upper and lower limits of $R_{\rm 
sh} = 4 R_{200}$ and $1.3 R_{200}$, respectively.

Figure~\ref{fig:time} shows that the electron-ion equipartition time 
$t_{ei}$ is always shorter than the shock elapsed time $t_{\rm sh}$ at 
radius $\lesssim 1.2 R_{200}\approx 90\arcmin$.  This suggests that 
electrons and ions should have enough time to share energy, and hence they 
should have the same temperature ($T_e=T_i$) inside that radius.  This 
is consistent with the simulations of \citet{WS09} 
that the non-equipartition effect 
should be small at radii $\lesssim R_{200}$, but only becomes important 
beyond that.  However, if the accretion shock radius is 
very small ($R_{\rm sh} < 1.3 R_{200}$) or internal accretion shocks 
develop close to $R_{200}$, a non-equipartition effect can be important 
near $R_{200}$.

The collisional ionization time $t_{\rm CIE}$ is 
about four to five times
longer than $t_{ei}$ near $R_{200}$.  For $R_{\rm sh} \lesssim 
1.3 R_{200}$, $t_{\rm CIE}$ is longer than $t_{\rm sh}$ at radii  
$\gtrsim R_{200} \approx 76\arcmin$.  
A non-equilibrium ionization (NEI) state is possible around $R_{200}$, and 
might be detected with high-resolution X-ray spectrometer \citep[see., 
e.g.,][]{WSJ11}.  However, if the accretion shock radius is 
larger ($R_{\rm sh} \gtrsim 1.3 R_{200}$), 
$t_{\rm CIE}$ is shorter than $t_{\rm 
sh}$, and therefore the gas is likely to be in CIE for radii $\lesssim 
R_{200}$.

\section{\lowercase{$f_{\rm gas,200}$}--$T_{500}$ and 
$K_{200}$--$T_{500}$ relations}
\label{sec:scaling}

Measuring the gas fraction of galaxy clusters and groups is of great 
interest.  It depends on the fraction that has converted to stars and 
the gas expelled by heating or during the cluster or group 
formation; thus it can be used to test structure formation theories.  
The gas fraction has also been used to constrain cosmological parameters 
\citep[e.g.,][]{ARS+08}.  X-ray studies of large samples of individual 
clusters and groups have been used to constrain the $f_{\rm gas}$--$T$ 
relations out to $R_{500}$ \citep[see, e.g.,][]{Sun12}.  Another 
important quantity is entropy, which reflects the thermodynamic history of 
hot gas.  When measuring the entropy at small radii $\lesssim 
0.15 R_{500}$, the $K_{\Delta}$--$T_{500}$ relations of groups and 
clusters have been found to deviate from the gravity heating-only 
baseline model, with a large scatter for low mass groups.  The scaling 
relations become consistent with the baseline model as the radii 
approach $R_{500}$ \citep{SVD+09,Pra+10}.  We extend earlier work to 
examine the $f_{\rm gas,200}$--$T_{500}$ and $K_{200}$--$T_{500}$ relations 
with the gas fraction and entropy measured at $R_{200}$.

We complied the enclosed gas-mass fractions ($f_{\rm gas, 200}$) and 
entropies ($K_{200}$) of galaxy groups and clusters measured out to 
$R_{200}$ using {\it Suzaku} data published in the literature.  One of 
the clusters, Virgo, which was measured with {\it XMM-Newton}, is also 
included to increase the number of data points in the sparse low mass 
group range.  The compiled results are listed in 
Table~\ref{table:sample}.  The scaled temperatures $T_{500}$ were taken 
from the work listed in column 3 of Table~\ref{table:sample}.  
The $T_{500}$ values reported are average temperatures measured from 
$0.15R_{500}$ to $R_{500}$ or close to this range.
The scaled radii $R_{200}$, $f_{\rm 
gas,200}$, $K_{200}$, and redshift $z$ were taken from the work listed
in column 8 of Table~\ref{table:sample}.  When there is no value at 
$R_{200}$ reported explicitly, we evaluated $f_{\rm gas,200}$ and 
$K_{200}$ at $R_{200}$ by interpolation or extrapolation in log space 
using the two nearest data points in the radial profile.  Extrapolations 
were done for two clusters with $R_{200}$ no more than $\sim$10\% beyond 
the outer boundary of the data.  The extrapolated values are enclosed in 
parentheses in Table~\ref{table:sample}.  

It would be ideal if the 
observations were taken in relaxed directions or azimuthally averaged 
over all directions to minimize biases along merging or filament 
directions.  However, most of the observations only covered one or a few 
narrow directions due to the expensive exposure required.  Therefore, to 
increase the sample size, we also included clusters with observed 
directions along merging or filament directions.  The 
data of the merging or filament directions used in our sample do not 
appear to introduce significant biases to the scaling relations (see 
Figures~\ref{fig:f-T} and \ref{fig:K-T} below). When there was 
more than one direction reported in the literature, we chose the 
relaxed direction if there was a significant difference between the relaxed and 
non-relaxed directions.  Otherwise, the azimuthal average quantities were
used.  The last column of Table~\ref{table:sample} describes 
the directions of the observations. The errors of all data have been 
converted to 90\% confidence range.

Figure~\ref{fig:f-T} shows the $f_{\rm gas,200}$--$T_{500}$ relations 
for a wide range of temperatures.  We fitted the data with the Bivariate 
Correlated Errors and Intrinsic Scatter (BCES) method \citep{AB96}.  The 
$(Y \mid X)$ regression was used because the temperature errors are 
generally smaller than the gas fraction errors.  If we consider only
those with total mass measured using the HSE method, the best-fit power law 
slope is $0.328 \pm 0.166$, suggesting an increasing $f_{\rm gas,200}$ 
as temperature increases.  At $R_{500}$, \citet{SVD+09} 
found that the slope is $\sim$$0.16$--$0.22$, while it can be as steep
as 0.32, as found by \citet{LRS15}.  The error of the slope at $R_{200}$ is too 
large to tell if there is any difference from those at smaller radii.  
If we include the three clusters with total masses measured using the weak 
lensing method, the best-fit power law slope decreases to $0.168 \pm 
0.221$, and hence the correlation becomes insignificant.  
Note that adding these three lensing clusters can introduce 
biases to the results, because the values of $f_{\rm gas,200}$
for other clusters might decrease with 
weak lensing mass measurements.
For example, if the ratio of HSE mass and weak lensing mass are the same for 
all groups or clusters, the $f_{\rm gas,200}$ slope of the HSE sample will
be the same as that of the weak lensing sample.
Note also that only two reliable weak lensing mass measurements are 
available in these data, and therefore our discussions related to weak 
lensing measurements can be strongly biased.

\begin{figure}
\includegraphics[width=0.4\textwidth, angle=270]{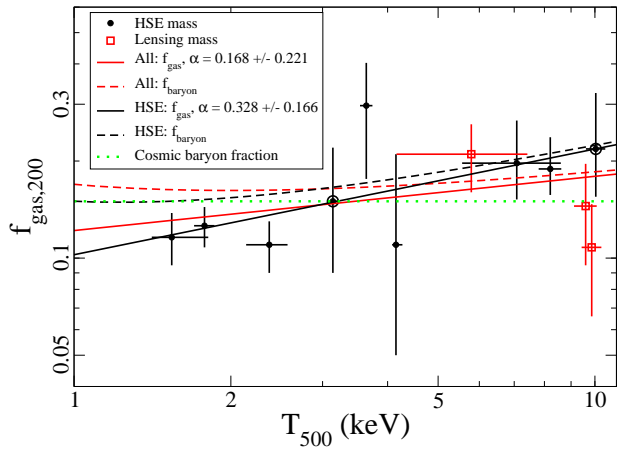}
\caption{Enclosed gas fraction within $R_{200}$ vs. 
$T_{500}$.  Black open circles
indicate clusters measured along merging or filament directions. The 
black solid line is the best fit to groups and clusters with masses 
measured using the HSE method.  The red solid line is the best fit 
to all groups and clusters in the sample (masses measured with either HSE 
or weak lensing methods).  The corresponding dashed lines are the total 
baryon fraction within $R_{200}$, calculated by adding the gas fraction 
with the stellar mass fraction.
Note that the small sample size of weak lensing mass measurements can 
introduce a strong bias.}
\label{fig:f-T}
\end{figure}

\begin{figure}
\includegraphics[width=0.4\textwidth, angle=270]{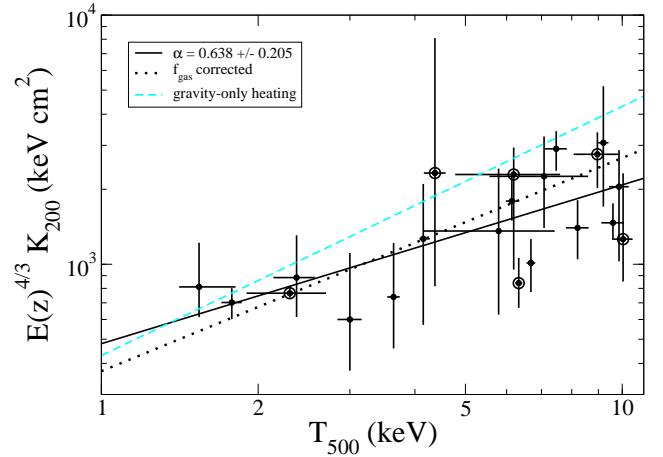}
\caption{Entropy at $R_{200}$ vs. $T_{500}$.  Black open circles 
indicate clusters measured along merging or filament directions.  The 
black solid line is the best fit to all groups and clusters in the 
sample.  The black dotted line is the gas corrected relation.  The cyan 
dashed line is the baseline model.}
\label{fig:K-T}
\end{figure}

\begin{deluxetable*}{ccccccccc}
\tabletypesize{\scriptsize}
\tablewidth{0pt}
\tablecolumns{5}
\tablecaption{Clusters with measurements out to $R_{200}$ 
with {\it Suzaku},
plus Virgo (measured with {\it XMM-Newton}).  
\label{table:sample}
}
\tablehead{
\colhead{Name} &
\colhead{$T_{500}$} &
\colhead{Ref.~1$^{\rm a}$} &
\colhead{$R_{200}$} &
\colhead{$f_{\rm gas, 200}$} &
\colhead{$K_{200}$$^{\rm b}$} &
\colhead{$z$} &
\colhead{Ref.~2} &
\colhead{Note$^{\rm c}$}
\\
\colhead{} &
\colhead{(keV)} &
\colhead{} &
\colhead{(Mpc/arcmin)} &
\colhead{} &
\colhead{(keV~cm$^2$)} &
\colhead{} &
\colhead{} &
\colhead{}
}
\startdata
Antlia        & $1.54^{+0.27}_{-0.13}$ & (6) & 0.887 / 76   & $ 0.116^{+0.022}_{-0.021}$  & $ 806^{+407}_{-197}   $ &0.00933 &  This work & Relaxed \\
RX J1159+5531 & $1.78^{+0.08}_{-0.08}$ & (5) &  0.871/ 9.7  & $ 0.126^{+0.018}_{-0.018}$ & $ 667^{+95}_{-95}     $ &0.081 &  \citet{SBG+15} & Full \\
Virgo         & $2.3^{+0.4}_{-0.4}$    & (6)$^{\star}$ &  1.08 / 234  & \nodata                     & $ 763^{+53}_{-31}     $ &0.00436 &  \citet{UWS+11} & Filament \\
ESO 3060170   & $2.37^{+0.20}_{-0.23}$ & (3) &  1.15 / 28   & $ 0.11^{+0.02}_{-0.02 }$    & $ 865^{+414}_{-264}   $ &0.0358 &  \citet{SWM13} & Relaxed \\
Centaurus     & $3.00^{+0.16}_{-0.16}$ & (6) &  1.130/ 84   & \nodata                    & ($ 596^{+507}_{-225}  $) &0.0109 &  \citet{WFS+13} & Relaxed \\
A1750N        & $3.14^{+0.08}_{-0.07}$ & (6) &  1.37 / 14.1 & $ 0.15^{+0.07}_{-0.06 }$    & \nodata                 &0.0832 &  \citet{Bul+16} & Merging \\
Hydra A       & $3.64^{+0.10}_{-0.10}$ & (2) &  1.183/ 17.8 & $ 0.297^{+0.106}_{-0.121}$  & $ 715^{+461}_{-271}   $ &0.0539 &  \citet{Sat+12} & Relaxed \\
A1750C        & $4.15^{+0.12}_{-0.07}$ & (6) &  1.57 / 16.2 & $ 0.11^{+0.10}_{-0.06 }$    & $1198^{+792}_{-657}   $ &0.0864 &  \citet{Bul+16} & Relaxed \\
A3376         & $4.37^{+0.21}_{-0.21}$ & (2) &  1.86 / 34.6 & \nodata                   & ($2258^{+5602}_{-1465} $) &0.046 &  \citet{ATN+12} & Merging \\
A1246$^{\dagger}$ & $5.79^{+1.63}_{-1.63}$ & (1) &  1.97 /10.3  & $ 0.21^{+0.05}_{-0.05 }$    & $1200^{+936}_{-646}   $ &0.1902 &  \citet{SMY+14} & Full \\
A85           & $6.19^{+1.41}_{-1.41}$ & (1) &  1.81 / 28.2 & \nodata                     & $2215^{+627}_{-1295}  $ &0.055 &  \citet{IWS+15} & Merging \\
A1795         & $6.14^{+0.16}_{-0.16}$ & (2) &  1.9  / 26   & \nodata                     & $1723^{+432}_{-285}   $ &0.063 &  \citet{Bau+09} & Relaxed? \\
A3667         & $6.33^{+0.10}_{-0.10}$ & (2) &  2.26 / 34.1 & \nodata                     & $ 812^{+213}_{-166}   $ &0.0556 &  \citet{AdK+12} & Merging \\
Perseus       & $6.68^{+0.13}_{-0.13}$ & (4) &   1.8 / 82   & \nodata                     & $1001^{+247}_{-236}   $ &0.0179 &  \citet{Urb+14} & Relaxed \\
A2029         & $7.08^{+1.52}_{-1.52}$ & (1) &  1.92 / 22.0 & $ 0.197^{+0.070}_{-0.045}$  & $2148^{+956}_{-814}   $ &0.0767 &  \citet{WFS+12a} & Full \\
A1413         & $7.47^{+0.36}_{-0.36}$ & (1) &  2.24 / 14.8 & \nodata                     & $2651^{+469}_{-488}   $ &0.1427 &  \citet{Hos+10} & Relaxed? \\
PKS 0745-191  & $8.21^{+0.41}_{-0.41}$ & (4) &  2.0  / 17.4 & $ 0.189^{+0.048}_{-0.032}$ & $1312^{+381}_{-331}   $ &0.1028 &  \citet{WFS+12b} & Full \\
A2744         & $8.96^{+0.89}_{-0.89}$ & (1) &  2.0  /  7.3 & \nodata                     & $2238^{+504}_{-599}   $ &0.308 &  \citet{IOA+14} & Filament \\
Coma          & $9.20^{+0.21}_{-0.21}$ & (4) &  2    / 70   & \nodata                     & $3031^{+2080}_{-1352} $ &0.0231 &  \citet{Sim+13} & Relaxed \\
A1835$^{\dagger\dagger}$ & $9.60^{+0.48}_{-0.48}$ & (1) &   2.21/ 9.08 & $ 0.145^{+0.051}_{-0.050}$  & $1238^{+244}_{-182}   $ &0.253 &  \citet{Ich+13} & Full \\
A1689$^{\dagger\dagger}$ & $9.86^{+0.43}_{-0.43}$ & (1) &  2.4  / 13   & $ 0.108^{+0.039}_{-0.042}$  & $1817^{+730}_{-906}   $ &0.1832 &  \citet{Kaw+10} & Full \\
A2142         & $10.04^{+0.43}_{-0.43}$& (2) &  2.48 / 24.8 & $ 0.218^{+0.107}_{-0.063}$  & $1191^{+997}_{-386}   $ &0.0909 &  \citet{AHI+11} & Filament
\enddata
\tablenotetext{a}{Reference of $T_{500}$ taken from: (1) \citet{MSF+15}, (2) \citet{Vik+09}, (3) \citet{SVD+09}, 
(4) \citet{DM02}, (5) \citet{HBB+12}, (6) same as column (8).}
\tablenotetext{b}{The values in parentheses are from extrapolation (see text).}
\tablenotetext{c}{Directions of the observations indicated by the authors in column (8).  
``Relaxed'' includes directions away from the merging or filament axis.
``Merging'' includes either direction along the merging axis.
``Full'' indicates average value using four or more directions.
The ``?'' symbol indicates that the condition is not explicitly written by the authors, 
but it is inferred by the content of the paper.}
\tablenotetext{${\star}$}{Error of $T_{500}$ was not reported, and we estimated it from the fluctuations in its temperature profile.}
\tablenotetext{${\dagger}$}{$f_{\rm gas}$ is measured using the weak lensing mass instead of HSE mass, where the weak lensing mass is estimated using a weak lensing template profile from a sample of low mass clusters rather than measured directly.}
\tablenotetext{${\dagger\dagger}$}{$f_{\rm gas}$ of the cluster/group calculated using the weak lensing mass instead of HSE mass.}
\\
\end{deluxetable*}

We estimated the enclosed baryon fraction $f_{\rm baryon}$ by adding the 
enclosed stellar fraction at $R_{200}$ measured by \citet{And10}.  The 
results are shown as dashed lines in Figure~\ref{fig:f-T}.  For the 
sample with only HSE mass measurements, $f_{\rm baryon}$ rises from 0.15 
for 1\,keV groups to 0.22 for 10\,keV clusters, but note that the 
uncertainly is too large to confirm the trend or to tell whether there 
is an apparent baryon excess for the massive clusters.  
The excess in $f_{\rm baryon}$ for massive clusters with HSE mass 
measurements is only marginally significant.  There is no strong 
evidence of significant clumping around $R_{200}$.
When including 
the clusters with lensing mass measurements, $f_{\rm baryon}$ is 
essentially flat.  
The baryon fractions of massive clusters are 
consistent with the cosmic value within 
$\sim$20\%, which is 
smaller than the 
uncertainty of the measurement.  Therefore, there is no evidence of 
missing baryons in all mass ranges.  

The $K_{200}$--$T_{500}$ relation is plotted in Figure~\ref{fig:K-T}.  
We have also plotted the baseline entropy model
using equations~(\ref{eq:entropy}) and (\ref{eq:K200}), 
where we assume 
$R_{200}/R_{500} = 1.5$ when converting $\bar{T}_{200}$ to $\bar{T}_{500}$.
Using the BCES method, the best-fit power law slope is 
$0.638 \pm 0.205$, 
which is significantly smaller than the baseline 
entropy model of 1.
The slope is also smaller than that of 
the observed $K_{500}$--$T_{500}$ relation, which was found to be 
$0.994\pm0.054$ (1$\sigma$) by \citet{SVD+09} and $0.92\pm0.24$ 
(1$\sigma$) by \citet{Pra+10}. For low temperature groups with $T_{500} 
\lesssim 2.5$~keV, $K_{200}$ is consistent with the baseline model.  
Above that, $K_{200}$ is significantly smaller.  While there are 
significant entropy excesses for groups at $R_{500}$ compared to the 
baseline model \citep{SVD+09}, we do not see any excess for groups at 
$R_{200}$.  This suggests that non-gravitational heating is important at 
radii $\lesssim R_{500}$ but not around $R_{200}$.  There is also some 
evidence of entropy excess in more massive clusters at $R_{500}$ shown 
by \citet{SVD+09}.  
However, Figure~\ref{fig:K-T} 
shows that the entropy 
at $R_{200}$ is clearly smaller than the baseline model.
Such an entropy deficit in massive clusters measured with {\it Suzaku} 
was discussed by various authors
(see Section~\ref{sec:dis2}), 
although \citet{EMV+13} did not find a significant deficit in 
their stacked {\it ROSAT} and {\it Planck} sample.
\citet{Oka+14} found that the entropy scaling relation has a slope 
that agrees with the baseline model near $\sim$$R_{200}$, which is steeper 
than our finding, although they also noted that the normalization is 
lower than the baseline model.  If we only take the three clusters 
(Hydra A, A1835, A1689) that were also used in their four clusters sample, 
the $K$--$T$ slope is closer to 1.  Thus, the slope found in \citet{Oka+14} 
might be overestimated due to their small sample size.

We corrected the best-fit entropy by introducing the same gas correction 
factor as in Section~\ref{sec:DPE}.  
We used $f_{\rm gas,200}$ from the HSE sample 
(black solid line in Figure~\ref{fig:f-T})
so that a larger correction can be made.
The gas-corrected 
entropy is shown as a dotted line of Figure~\ref{fig:K-T}.  The corrected 
entropy has a slope of $0.86\pm0.23$, but the uncertainty is too large 
to tell whether it deviates from 1
of the baseline model.  The 
corrected entropy for massive clusters is still lower than the baseline 
model, suggesting that the entropy deficit at $R_{200}$ is not fully 
correlated to the gas fraction, unlike those connections found at smaller 
radii \citep{Pra+10}.  Thus, the entropy deficits for massive clusters 
might require biases or deviations in temperature measurements.

\section{Discussion}
\label{sec:discussions}

\subsection{Diversity of Groups Out to R$_{200}$}
\label{sec:dis1}

The Virgo Cluster (or group) is a cool core cluster with a similar 
temperature \citep[2.3\,keV;][]{UWS+11} and 
mass \citep[$M_{500}$\,$\approx$\,$10^{14} M_{\odot}$:][]{WFS+13} as Antlia.  Virgo is the nearest cluster 
and Antlia is the nearest non-cool core cluster, making them an ideal 
pair to compare the difference between the cool core and non-cool core 
groups out to the virial regions with the best spatial resolution.  Note 
that the thermodynamic properties of Virgo were also studied in one 
direction (north) out to $\sim$$R_{200}$ by \citet{UWS+11}.
Virgo was studied along its major axis in X-rays where gas could be 
accreted faster, while Antlia was studied between the major and minor 
axes.

Besides the fluctuations of temperature due to small-scale structures 
resolved with {\it XMM-Newton} \citep{UWS+11}, we found that both Antlia 
and Virgo follow the same average scaled temperature profile of many 
clusters compiled by \citet{RBE+13} between $\sim$$0.1$ and $1 R_{200}$.  
The pressure profile of Virgo shows significant fluctuations, 
but the general trend appears to be flatter and slightly higher than the 
universal pressure profile beyond $\sim$$0.4 R_{200}$ \citep[Figure~6 
in][]{UWS+11}, in contrast to Antlia where the universal pressure 
profile is closely followed.  Thus, the density slope is flatter than 
that of Antlia as $n \propto P / T$, which can be seen in 
Figure~\ref{fig:npeprofile}.  The entropy profile of Virgo is also 
flatter as $K \propto T/n^{2/3}$.  Figure~10 in \citet{WFS+13} shows 
that, in spite of the large fluctuations, the entropy of Virgo is almost 
constant beyond $\sim$$0.3 R_{200}$.  Similar to Antlia, the entropy
beyond $\sim$$R_{500}$ approaches the gravity heating-only model.  
There appears to be some weak evidence that the entropy might be 
dropping slightly around $R_{200}$, but the fluctuations are too large 
to be conclusive.  Note that the last data point in the density in Figure~6 
of \citet{UWS+11} is likely biased high by about a factor of two due to 
the 
ringing effect in deprojection (see text in their paper),
and thus the 
entropy in Figures~10 and 12 of \citet{WFS+13} is biased low by about a 
factor of $2^{2/3} = 1.6$.  Taking this into account eliminates the 
entropy deficit at $R_{200}$ in Virgo, as well as the claim of clumping 
based on entropy deficit or self-similar considerations \citep{WFS+13}.  
Thus, non-gravitational effects in the Virgo outskirts are probably not 
very strong, similar to Antlia.  The flatter and stronger entropy excess 
inside $\sim$$R_{500}$ in Virgo suggests a stronger heating compared to 
Antlia, perhaps due to the strong AGN feedback from the supermassive 
black hole in M87.

We also compared the dynamically young Antlia to the dynamically evolved 
fossil groups that ESO 3060170 studied along one direction \citep{SWM13} and 
RX J1159+5531 studied in all directions \citep{HBB+12, SBG+15}.  RX 
J1159+5531 ($M_{500} \sim 6\times 10^{13} M_{\odot}$, $z=0.081$) has a 
similar total mass as Antlia, while ESO 3060170 ($M_{500} \sim 10^{14} 
M_{\odot}$, $z=0.0358$) is slightly more massive.

The entropy profiles are quite similar in the sense that they all rise 
from the center all the way out to $R_{200}$, and the values at 
$R_{200}$ are consistent with the gravity heating-only predictions
(see also Figure~\ref{fig:K-T}).  
The entropy slope of RX J1159+5531 in the outer regions between 
$\sim$0.1--$1R_{200}$ is also quite flat, with a power law index of 
about 0.5, lower but still consistent with that of 
$0.69^{+0.22}_{-0.24}$ for Antlia.  For ESO 3060170, there are some 
indications that the entropy near $R_{200}$ drops slightly but not 
significantly 
if the ringing effect is corrected, 
and thus it is consistent with the general trend of Antlia 
and RX J1159+5531. The density power law index of RX J1159+5531 is about 
$\alpha \sim 1.4$, which is flatter than Antlia of $\alpha = 
1.75^{+0.27}_{-0.24}$.  That of ESO 3060170 is much steeper, with 
$\alpha \approx 2.3$, closer to those of the massive clusters.  The gas 
($\sim$baryon) fractions of all groups are consistent with the cosmic 
value out to the virial regions, showing no evidence of strong clumping.  
While both Antlia and ESO 3060170 follow the universal pressure profile, 
the pressure of RX J1159+5531 is significantly higher beyond $\sim$$0.5 
R_{500}$, and deviates by more than a factor of two beyond 
$\sim$$R_{500}$.  Note that the universal pressure profile has a rather 
small scatter of $\sim$$30\%$ between 0.2 and $1R_{500}$ for massive 
clusters and a similar scatter for groups \citep{APP+10,SSV+11}.  
Compared to density or entropy, 
the universal pressure profile is believed to be less sensitive to 
dynamical history and non-gravitational physics.
One might expect that 
mergers or unrelaxed clusters or groups could have such larger deviations 
in their pressure profiles, but comparing the dynamically old 
RX J1159+5531 with large 
deviations with the dynamically young Antlia shows the opposite trend.  
Thus, RX J1159+5531 might be an exception.

In summary, we observe a diversity of ICM properties for different low 
mass groups.  
Strong disturbances such as cold fronts or sloshing can induce 
significant deviations from the universal pressure and baseline 
entropy profiles \citep[e.g.,][]{Urb+14}.
The dynamically young Antlia is surprisingly relaxed in 
the sense that it follows the universal pressure profile closely and the 
entropy profile approaches the gravity heating-only model out to 
$\sim$$R_{200}$. Furthermore, no significant fluctuations in temperature 
and density (or pressure and entropy) profiles were found.
The 
dynamically older cool core Virgo shows evidence of strong heating, 
perhaps by AGN feedback, which might affect the entropy and pressure out 
to $\sim$$R_{500}$.  One of the dynamically old fossil groups, ESO 
3060170, appears to be relaxed, but the other fossil group RX J1159+5531 
surprisingly shows a strong deviation from the universal pressure 
profile.  There is no strong evidence of entropy dropping and also no 
evidence of clumping in all these systems near the viral regions 
$\sim$$R_{200}$, but higher quality data and 
broader azimuthal coverage 
is needed to confirm these results for Antlia, RX 
J1159+5531, and Virgo.

\subsection{Comparison to massive clusters and implications for physics 
in cluster outskirts}
\label{sec:dis2}

More than a dozen clusters have been studied with {\it Suzaku} in detail 
out to $\sim$$R_{200}$ \citep[see a review of][]{RBE+13}.  Most of these 
are massive ($T > 3$\,keV) clusters, and we compare some surprising 
results found in these massive clusters with Antlia and other groups.

In Figure~10 of \citet{WFS+13}, the entropy near $R_{200}$ for massive 
clusters ($M_{500} >$ a few $10^{14} M_{\odot}$) is significantly smaller 
than the gravity heating-only 
baseline model, while the others are less clear.  
For the lowest mass groups, RX J1159+5531 and Virgo in their sample
(also the UGC\,03957 group  recently studied by \citealt{TLR+16}), 
the entropy profiles are consistent with 
the baseline model near 
$R_{200}$, similar to Antlia (Section~\ref{sec:dis1}).
Note that the {\it Suzaku} results by \citet{WFS+13} are in some tension 
with the joint {\it ROSAT} and {\it Planck} data analysis of massive clusters
by \citet{EMV+13}, with the latter suggesting a continuously 
increasing entropy out to $\sim$$R_{200}$ with only a minor (although still
notable) entropy deficit at large radii.  
\citet{FL14} argued that the discrepancy is due to 
the pressure (instead of temperature) profiles used by \citet{EMV+13} to 
determine the entropy, while pressure is insensitive to entropy and 
therefore temperature should be used instead.
If the entropy deficit around $R_{200}$ for 
massive clusters is real 
as measured by \citet{WFS+13}, 
this will imply that some non-gravitational 
processes are 
responsible for altering the entropy of massive cluster outskirts.  
For low mass groups 
where their entropy profiles follow the baseline model at large radii, 
perhaps non-gravitational processes are not important in their outskirts 
or different non-gravitational effects might cancel out
to bring back the final entropy profiles to the baseline model.

Another surprising result is that the measured gas ($\sim$baryon) 
fractions of some massive clusters appear to be significantly higher 
than the cosmic value \citep[e.g.,][]{Sim+11}, suggesting biases in the 
measured gas density (mass) or total mass;  
although Figure~\ref{fig:f-T} shows that the excess in massive clusters is 
only marginally significant.
However, for the low mass 
groups discussed in Sections~\ref{sec:dis1}, there is no evidence for any 
excess of gas fraction, suggesting that the bias is perhaps minimal on 
group scales.

One of the explanations for the entropy and gas fraction deviations is 
the clumping or inhomogeneity in gas \citep{Sim+11}, which is expected 
to be present at some level.  The idea is that the emission measure is $\propto 
n_e^2$, and therefore we are measuring the average $\langle n_e^2 
\rangle$, which is always higher than $\langle n_e \rangle^2$ for clumpy 
gas.  Thus, the measured electron density of clumpy gas is biased high 
and the entropy ($K \propto T/n_e^{2/3}$) is biased low.  If the clumps 
are cool, this can further bias low the entropy. 
Numerical simulations 
predict that clumping is stronger for more massive clusters above $10^{14} 
M_{\odot}$ beyond $\sim$$R_{200}$ \citep{NL11}.  
However, in these 
simulations, there appears to be no difference in the clumping level for 
low or high mass clusters
{\it inside} 
the regions of $\sim$$R_{200}$.  
This is inconsistent with the observations that more clumping is seen
{\it inside}
$R_{200}$ for more massive clusters.
If the low mass 
groups are indeed as clumpy as the massive clusters inside $R_{200}$ as 
predicted, this implies that the true average gas density is lower, and hence 
the true entropy is higher than the baseline model.  More theoretical 
work on the degree of clumpiness in groups is needed to address this issue.

\citet{WFS+13} attempted to separate whether the entropy deficits are 
due to the deviation of temperature or density from self-similar models.  
By assuming the self-similar entropy and the universal pressure 
profiles as the baseline models of the ICM, they suggest that the biases 
inside $\sim$$R_{200}$ are primarily due to the bias in density, while 
beyond that they are due to both biases in temperature and density. In fact, 
some clusters do not show bias in density but in temperature.  
Our results of the $K_{200}$--$T_{500}$ relation also 
suggest that bias in temperature is needed.
This 
might suggest that different mechanisms are working at different radii 
and also in different environments.  For example, clumping, which may 
not bias temperature \citep[see the discussion in, e.g.,][]{WFS+13}, 
might start to be important beyond $R_{500}$.  Beyond $R_{200}$, 
perhaps, e.g., non-equipartition electrons and ions might bias low the 
temperature, and NEI might bias high the electron density 
and bias low the temperature \citep[][see below also]{RBE+13}.  
Multi-phrase gas might also bias low the temperature.
In particular, in Figure~13 of \citet{WFS+13}, the temperature is 
strongly biased low beyond $R_{200}$, which drops to about 0.1--0.2 of 
the self-similar model, suggesting the need for some mechanisms that can 
strongly affect the temperature.

Mass determination using weak lensing has provided important insights to 
the origins of the gas fraction bias and entropy deficit in some massive 
clusters \citep{Kaw+10,Ich+13,Oka+14}.  For Hydra A, Abell 1689, Abell 
1835, and Abell 478, where the first three are in the sample of 
\citet{WFS+13}, the HSE masses determined from X-ray are significantly 
smaller than the weak lensing masses near $R_{200}$.  
When using the 
lensing masses instead of the HSE masses to calculate the gas fractions, 
they approach the cosmic value near $R_{200}$, indicating that the 
gas fraction bias is mainly due to a bias in the HSE mass rather than
a bias in the gas density determination due to clumpy gas.
This is consistent with our results
on the $f_{\rm gas,200}$--$T_{500}$ relations
in Section~\ref{sec:scaling},
although the uncertainties of the relations are still quite large.
\citet{Oka+14} also argue that the entropy 
deficit near $R_{200}$ is primarily due to the steepening in temperature 
measurements, rather than a shallower density slope.  Thus, the joint weak 
lensing and X-ray studies suggest that the breakdown of HSE is more 
important than clumping in cluster outskirts \citep{Ich+13}.  
Additional 
pressure support from turbulence or bulk motions 
\citep[e.g.,][]{LKN09}, higher ion temperature due to non-equipartition, 
and cosmic rays \citep[e.g.,][]{LdK10,VBG+12} may be responsible for the 
non-HSE effects.  Weaker accretion shocks have also been suggested to be 
responsible for the steepening in the outer temperature \citep{LFC10}.  For 
groups, there is also some 
indication
that HSE masses are significantly 
lower than weak leasing masses \citep{Ket+13}.  Thus, energy in addition 
to those inferred by thermal electrons should be present.  Because we do 
not see an entropy deficit in groups, the total energy injected to the gas 
might be higher than the pure gravity heating near $R_{200}$.  The true 
gas ($\sim$baryon) fraction might in fact be lower than the cosmic 
value, suggesting that gas might be pushed away by the higher energy 
injection.  It would be interesting to study whether previous AGN 
feedback can inject enough energy to affect the gas properties out to 
$R_{200}$ in groups \citep[e.g.,][]{Fuj01}.

As mentioned, non-equilibrium effects due to low density plasma have 
been considered as another possible mechanism to explain the entropy 
deficit near $R_{200}$. The electron temperature behind a shock can be 
lower than the ion temperature, leaving a non-equipartition state of 
electrons and ions \citep[e.g.,][]{WS09,ANL+15}.  The ions can also be 
underionized after a shock, and this NEI plasma has a higher emissivity 
than the CIE plasma.  The soft X-ray emission between 0.3 and 1\,keV can 
be an order of magnitude higher for NEI plasma near the shock region 
\citep{WSJ11}.  One major uncertainty of whether these non-equilibrium 
processes take effect around $R_{200}$ is the location of accretion 
shocks where the effects are strongest.  By using spherical symmetric 
hydrodynamic simulations, \citet{WS09} show that the non-equipartition 
effect is at most a few percent near $R_{200}$ and stronger beyond that, 
while \citet{ANL+15} show that the effect can be up to $\sim$10\% at 
$\sim$$R_{200}$ using realistic 3D simulations.\footnote{Note that 
\citet{ANL+15} use a scale radius $R_{200m}$ defined according to the 
density of matter instead of the critical density, where $R_{200} 
\approx 0.6 R_{200m}$.}  Thus, clusters in a realistic environment might 
induce shocks at smaller radii, making the non-equipartition effect 
stronger at $R_{200}$. 

For a $\sim$$10^{15} M_{\odot}$ massive cluster, 
the NEI effect can bias high the soft X-ray emission by more than 10\% 
for regions where the non-equipartition effect starts to be important 
\citep{WSJ11}, and this can bias high the density measured by assuming 
CIE.  
The excess soft emission can also bias low the 
measured temperature \citep{RBE+13}.
Realistic 3D simulations
would be needed to test whether the NEI 
effect can introduce significant emission bias around 
$R_{200}$.\footnote{NEI was not included in the 
simulations by \citet{ANL+15}.} 
Because
the non-equipartition electron temperature is lower than the average 
temperature of the plasma, the HSE mass estimated using the electron 
temperature determined by spectral fitting can also be biased low by 
$\sim$$10\%$ near $R_{200}$ for massive clusters \citep{ANL+15}, and 
thus the gas fraction is further biased high.  The 3D simulations also 
predict that the non-equipartition effect is reduced along filaments, 
which is qualitatively consistent with the higher temperature along the filament 
directions in Abell 1689 \citep{Kaw+10} and Perseus \citep{Urb+14}.  
Moreover, the density bias is higher along the minor axis of Perseus, 
which is also qualitatively consistent with the NEI effect 
being stronger there, but in contrast to the clumping prediction where its 
effect is more significant along the filament (major) direction.  
Numerical simulations also suggest that these collisionless effects are 
more significant for more massive clusters.  For groups cooler than 
2--3\,keV, the effects near $R_{200}$ are negligible \citep{WS09, 
WSJ11,ANL+15}. Thus, both non-equipartition and NEI effects can 
potentially explain the entropy deficit and high-gas fraction near 
$R_{200}$ for massive clusters, and at the same time allow the normal 
entropy and gas fraction for low mass groups.

With current instruments, it is indeed not easy to distinguish the 
non-equipartition and NEI effects from other models, such as cooler 
clumps/subgroups or multi-temperature structures, where some or all of 
them can be working together.  Current support mainly comes from 
timescale estimates, suggesting that these collisionless effects are 
possible near $R_{200}$.  \citet{Hos+10} and \citet{AHI+11} estimated 
that the electron-ion equipartition timescales could be longer than the 
shock-elapsed timescales in the massive clusters Abell 1413 and Abell 
2142, respectively, suggesting possible non-equipartition of electrons 
and ions near $R_{200}$.  However, they assumed that the shock radii are 
at $R_{200}$, which is probably an underestimation for virial shocks, 
and thus might overestimate the significance of the non-equipartition 
effect.  For the massive merger Abell 3667, a shock was detected 
near $R_{200}$ and the timescale estimations indicate that it is 
possibly in non-equipartition and NEI \citep{FSN+10,ATN+12}.  Thus, 
merging might enhance these non-equilibrium effects near $R_{200}$ as 
expected. Our estimations for Antlia suggest that the gas around 
$R_{200}$ is probably in equipartition ($T_e = T_i$). 
It is also probably in CIE unless $R_{\rm sh} 
\lesssim 1.3 R_{200}$.  This is consistent with the predictions that 
these non-equilibrium effects should be small in groups
\citep{WSJ11,RBE+13}.  A more quantitative study will be needed to test 
whether or how much these non-equilibrium effects are responsible for 
the entropy and density biases in massive clusters. In the future, the 
most direct way to study these effects will be to measure the ion temperature 
by line width and ionization state by line ratio, which could be 
possible with the {\it Athena} mission \citep{WSJ11,Nan+13}.


\section{Summary and Conclusions}
\label{sec:summary}

We have presented {\it Suzaku} observations
of the nearest non-cool core cluster, Antlia, out to $1.3 
R_{200}$ in the east direction, which is between the major and minor 
axes of the X-ray emission
and also away from the large-scale filament direction.
{\it Chandra} and {\it XMM-Newton} data were 
used to minimize the point source contamination in all the {\it Suzaku} 
pointings.  {\it ROSAT} data were also used to ensure the consistency in 
the soft X-ray background determination.  Different systematic 
uncertainties were taken into account to ensure the results are robust.

The temperature of Antlia drops from about 2 keV near the center down to 
about 0.7 keV near $R_{200}$, which is consistent with many other clusters.  The 
projected temperature profile of Antlia is consistent with the average 
scaled profile of other groups and clusters out to $R_{200}$.  The power 
law index of the density profile beyond $\sim$$0.1 R_{200}$ is $\alpha = 
1.75^{+0.27}_{-0.24}$, which is significantly steeper than that of the cool core 
Virgo Cluster ($\approx 1.2$), but shallower than those of the massive clusters 
($\approx 2$--3).  The pressure of Antlia follows the universal 
profile out to $\sim$$R_{200}$.

The entropy profile increases all the way out to $\sim$$R_{200}$, with 
its value approaching the gravity heating-only baseline model, but a 
flatter power law index of $0.69^{+0.22}_{-0.24}$ compared to the 
baseline model of 1.1.  Thus, no entropy deficit is seen near $R_{200}$ 
as compared to some massive clusters.  The entropy inside $R_{500}$ is 
significantly higher than the baseline model, as has been found in many 
other groups.  Thus, some non-gravitational processes are responsible for 
the high central entropy.

The gas-mass fraction increases from the center and approaches the 
cosmic value near $1.3 R_{200}$.  Therefore, clumping is not significant 
in this direction.
 
The electron-ion equipartition timescale is shorter than the 
shock-elapsed timescale inside $R_{200}$, suggesting that electrons and ions 
inside this region should share the same temperature in this low mass 
group.  
Although the collisional ionization timescale can be much longer, 
the plasma in Antlia should still be in CIE near $R_{200}$ unless
its shock radius is smaller than about $1.3 R_{200}$.

We compiled X-ray measurements primarily using {\it Suzaku} 
observations in the literature.  The $f_{\rm gas,200}$--$T_{500}$ 
relation has a power law slope of $0.328 \pm 0.166$ for the sample with 
HSE mass measurements.  
After correcting 
for the stellar mass fraction, the enclosed baryon fraction at $R_{200}$ 
is consistent with the cosmic value.

The power law slope of the $K_{200}$--$T_{500}$ relation is $0.638 \pm 
0.205$, which is significantly smaller than the gravitation heating-only model, 
and somewhat smaller than the $K_{500}$--$T_{500}$ relation measured by 
\citet{SVD+09} or \citet{Pra+10}.  The gas corrected 
$K_{200}$--$T_{500}$ relation has a larger slope of $0.86\pm0.23$.  The 
corrected entropy for massive clusters is still lower than the baseline 
model.  Thus, the entropy deficit at $R_{200}$ is not fully accounted by 
the bias or deviation in the gas fraction, 
in contrast to the entropy deficit at smaller radii 
\citep{Pra+10}.

We compared the non-cool core Antlia with three other low mass groups, 
Virgo, ESO 3060170, and RX J1159+5531, out to $R_{200}$.  
Counterintuitively, the dynamically youngest Antlia is surprisingly 
relaxed compared with some other dynamically older groups.  
Observations 
in other directions of Antlia are needed to test whether this is due to 
azimuthal variation.
The dynamically older cool core Virgo appears to 
be strongly heated to a very high entropy out to $\sim$$R_{500}$, 
presumably by the strong AGN feedback.  While one of the dynamically 
evolved fossil groups, ESO 3060170, appears to be relaxed, the other 
fossil group, RX J1159+5531 (with full azimuthal coverage), deviates 
significantly from the universal pressure profile.  Thus, we observe a 
diversity of ICM properties for different low mass groups.

While massive clusters sometimes show an entropy deficit and an excess in gas 
fraction near $R_{200}$, there is no such evidence for the lower mass 
groups we considered.  This suggests that clumping and other 
non-equilibrium processes in low mass groups might not be as significant
as in the high mass systems.  We argued that current data are not sensitive 
enough to distinguish or address the contributions of different models 
to explain the entropy and gas fraction deviations in massive cluster 
outskirts.  More observations spanning a wide range of mass and more 
complete azimuthal coverage, as well as more theoretical efforts, are 
needed to understand the outskirts of galaxy clusters and groups.  
A direct detection of clumps and non-equilibrium effects in cluster 
outskirts may be possible with future missions, such as the {\it 
SMART-X} version of the {\it X-ray Surveyor} mission concept and 
the {\it Athena} observatory \citep{Vik+12,Nan+13}.

\acknowledgments

We thank Lucas Johnson, Dacheng Lin, Peter Maksym, Eric Miller, Evan 
Million, Yuanyuan Su, and Mihoko Yukita for useful discussions.
We thank all PIs of the relevant observations and 
authors of the published works we used for their original efforts.
KWW and JAI were supported by NASA ADAP grants 
NNX13AI52G and NNX13AI53G, as well as 
{\it Chandra} grant GO3-14128A.  
CLS was partially supported by {\it Chandra} grant GO5-16131X and NASA 
{\it XMM-Newton} grant NNX15AG26.
YF was supported by KAKENHI No. 15K05080.
THR acknowledges support by the Deutsche Forschungsgemeinschaft (DFG) 
through Heisenberg research grant RE 1462/5 and grant RE 1462/6.

\appendix

\section{A. Emission Measure and Density Deprojection}
\label{sec:EM}

\begin{figure}
\includegraphics[width=0.4\textwidth,angle=270]{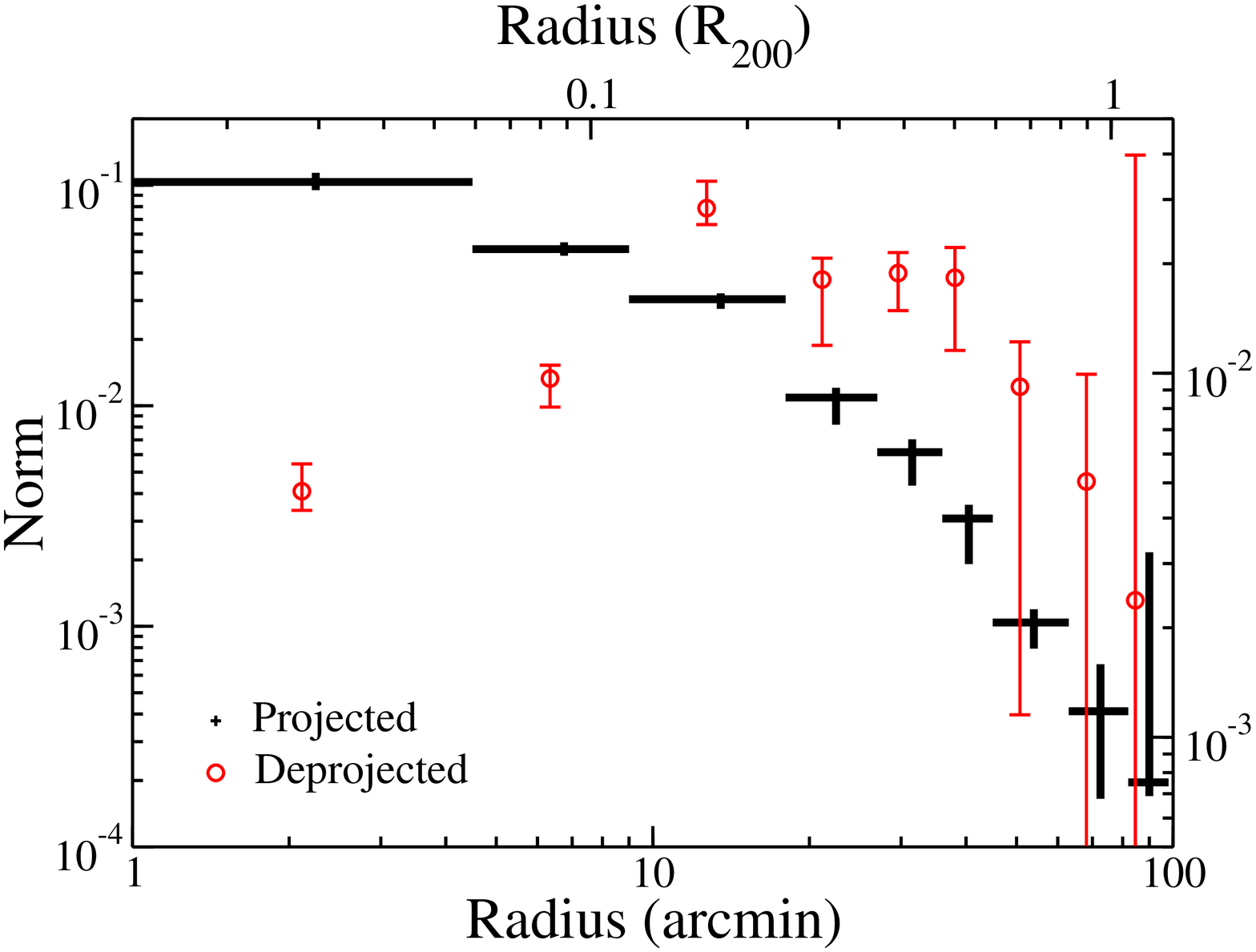}
\includegraphics[width=0.4\textwidth, angle=270]{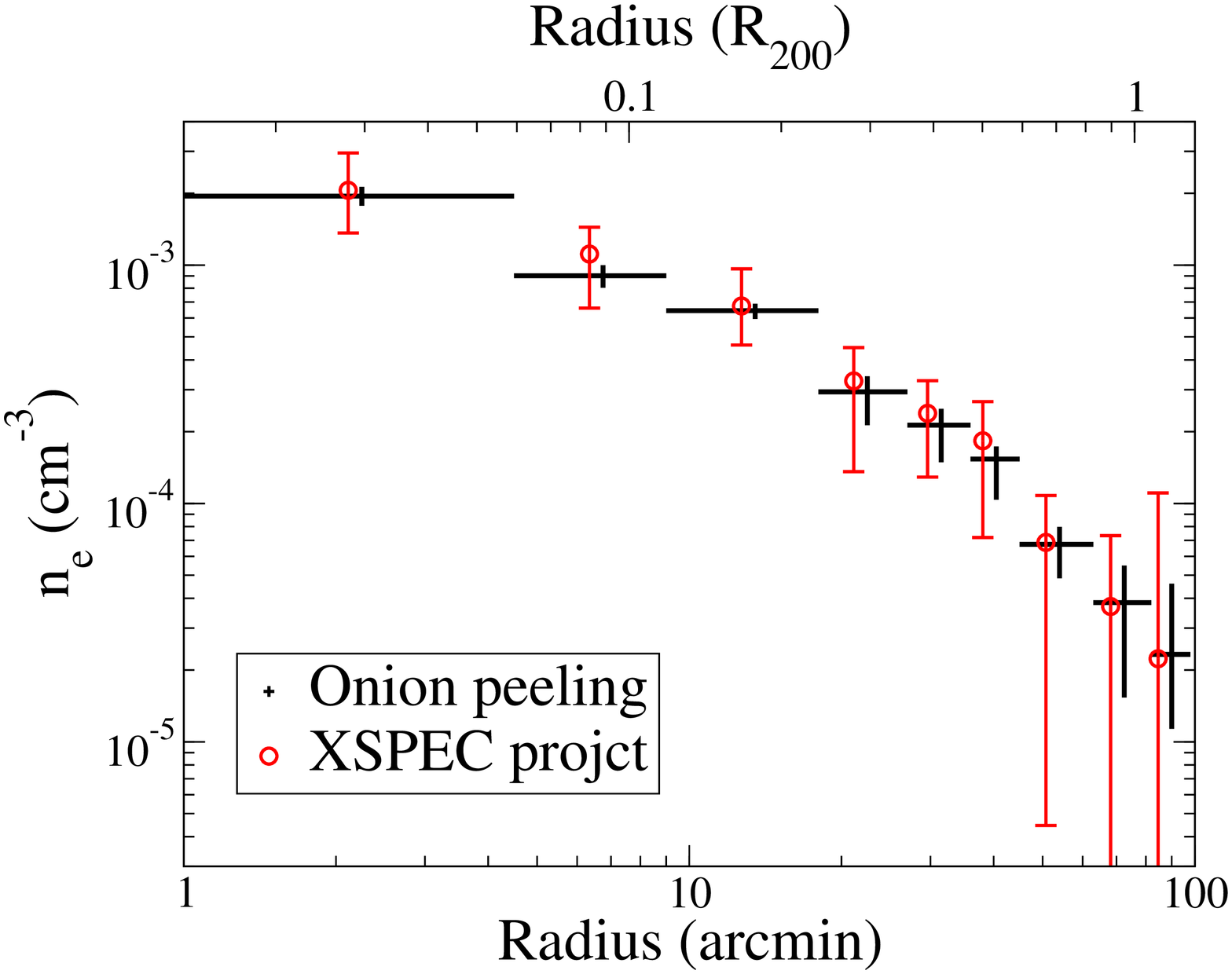}
\caption{Left panel: {\tt APEC} normalization per unit surface area
of the Antlia Cluster for the 
projected spectral analysis defined as 
$\frac{20^2 \pi}{\Omega}\frac{10^{-14}}{4\pi [D_A(1+z)]^2} \int n_e n_{\rm H} 
dV$~cm$^{-5}$~arcmin$^{-2}$, where $\Omega$ is the solid angle of the 
source spectral region in units of ${\rm arcmin}^2$
(black crosses corresponding to the left y-axis). 
Deprojected {\tt APEC} normalizations for the {\tt PROJCT} model are 
also plotted in red circles corresponding to the right y-axis and have 
been slightly shifted to the left for clarity. The deprojected {\tt 
APEC} normalizations are defined as $\frac{10^{-14}}{4\pi [D_A(1+z)]^2} 
\int n_e n_{\rm H} dV$~cm$^{-5}$, with the volume $V$ integrated over 
the spherical shell of the model.
Right panel: electron density profiles of Antlia using the onion peeling
(black crosses) and the {\tt XSPEC PROJCT} (red circles) methods.  The
red circle data points are slightly shifted to the left for clarity.}
\label{fig:surnorm}
\end{figure}

The left panel of Figure~\ref{fig:surnorm} shows the {\tt XSPEC} {\tt 
APEC} normalization
per unit surface area of the ICM for the projected spectral analysis, 
which is proportional to the emission 
measure of $\int n_e^2 dl$, where $n_e$ is the electron density and $l$ 
is the column length along the line of sight.  Systematic uncertainties 
have been added in quadrature to the statistical uncertainties.  The 
normalization per unit surface area drops from the center out to 
$\sim$$R_{200}$.  
The uncertainty of the last data point is too large to tell whether the 
emission measure continues to drop beyond $R_{200}$.

We also plot the deprojected {\tt APEC} normalization for the {\tt 
PROJCT} model in the left panel of Figure~\ref{fig:surnorm}, and this 
normalization is
proportional to the emission integral $\int n_e^2 dV$, where $V$ is the 
volume of the full spherical shell assumed in the deprojection.  As can 
be seen in Appendix~\ref{sec:app1}, the deprojected normalizations are 
subject to very large systematic uncertainties in the outer regions, 
and we can basically only constrain the upper limits of the outer two 
bins.  Nevertheless, the deprojected norms can be used to check for 
consistency when deriving the electron density using both 
the projected and deprojected spectral normalizations.

With the projected normalizations (or emission measure) 
determined in each annulus and assuming spherical symmetry, we can 
deproject the density profile using the onion peeling method outlined in 
\citet{KCC83} or \citet{WSB+08}.  In brief, this technique calculates 
the emission of each spherical shell starting from the outermost annulus 
toward the center, and the emission measure of each subsequent shell is 
calculated by subtracting the projected emission measure from the outer 
shells.

In doing the onion peeling deprojection, ignoring X-ray emission outside 
the last data bin can bias high the density at the last few data bins, 
similar to the bias seen in \citet{UWS+11}.  We corrected for this edge 
effect by extrapolating the normalization profile out to 150\arcmin\ 
using a power law fit to the data between 27\arcmin\ and 98\arcmin.  The 
errors of the extrapolated data were determined from the errors of the 
power law fit. We made $10^6$ simulated emission measure profiles for 
deprojections. The median of the electron density profile of 
Antlia is shown in the right panel of Figure~\ref{fig:surnorm}.  
The errors were 
estimated by the $10^6$ Monte Carlo simulations.

The deprojected {\tt APEC} normalizations of the {\tt PROJCT} model were 
directly converted to electron density (red circles in the right panel 
of Figure~\ref{fig:surnorm}).  The uncertainties of the density using 
the
{\tt PROJCT} model were much larger than the onion peeling method using 
the projected spectral normalizations; in particular, the outer few data 
bins are essentially unconstrained.  Nevertheless, the inner regions are 
consistent with one another, suggesting that the results are reliable.  
We can therefore take the average of the density profiles derived by the 
two methods using the standard weighted mean $\mu = \sigma \sum 
x_i/\sigma_i^2$, where the index $i$ indicates the method used, $\sigma 
= (\sum 1/\sigma_i^2)^{-1}$ is the error on the weighted mean, and $x_i$ 
and $\sigma_i$ are the density and its error (90\% confidence region 
including both statistical and systematic errors) of the associated 
method, respectively.  The average density profile is shown in the upper
panel of Figure~\ref{fig:npeprofile}.  This average density profile was 
used in all the calculations.

\section{B. Systematic Uncertainties}
\label{sec:app1}

\begin{figure*}
\includegraphics[width=0.82\textwidth, angle=270]{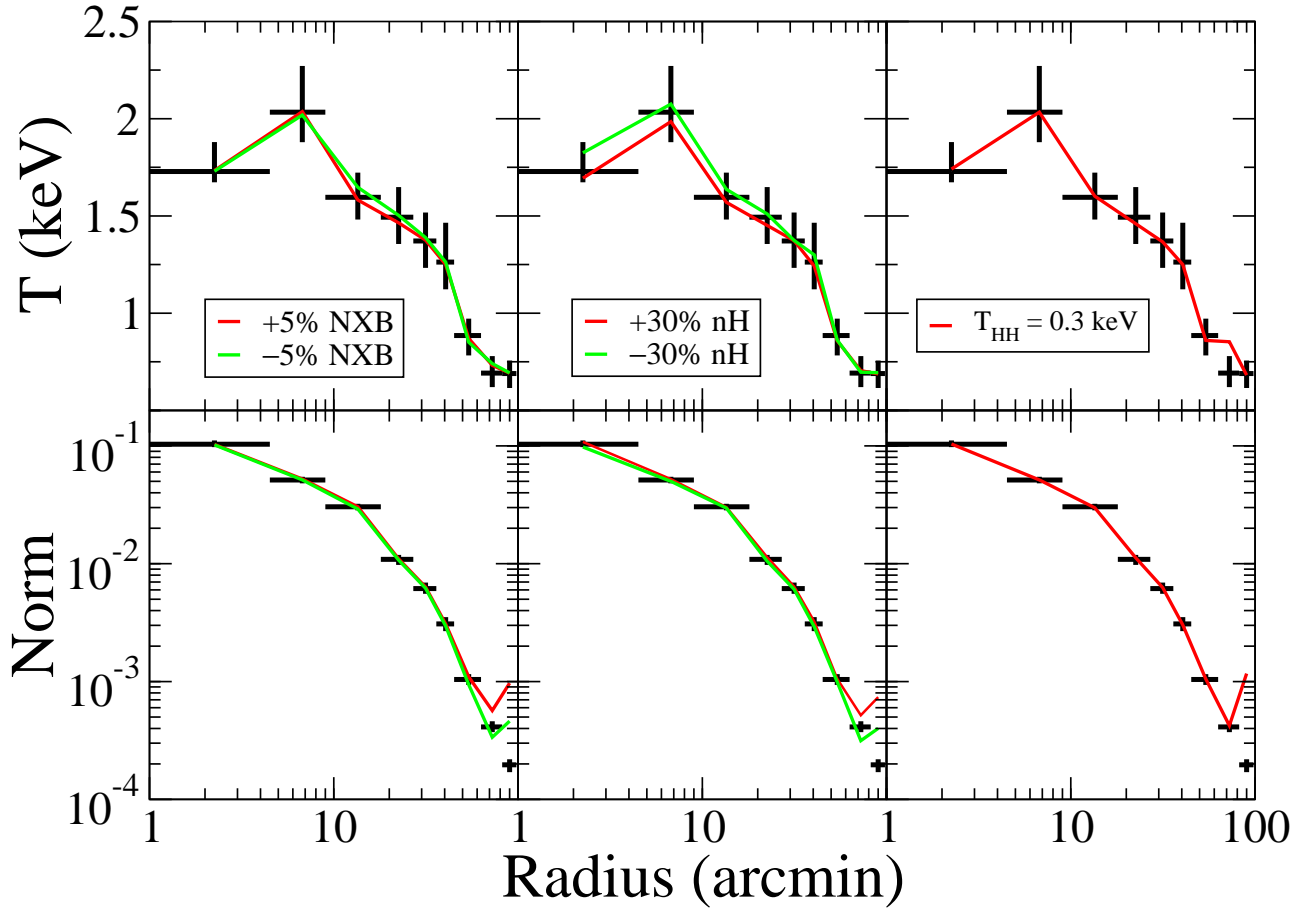}
\caption{
Left column: systematic uncertainties introduced by NXB uncertainties.  
The upper panel shows the temperature profile of the 
nominal single temperature model (black).
Model with NXB contribution 
increased (decreased) by 5\% is shown in red (green).  The lower panel 
shows the corresponding {\tt APEC} normalization per unit area.
Middle column: 
similar to left column, but with the red (green) lines representing a model with 
Galactic absorption ($n_{\rm H}$) increased (decreased) by 30\%.
Right column: similar to left column, but with the red line representing 
the model with the Galactic hot halo temperature fixed at $T_{\rm HH}=0.3$\,keV. 
For all 
panels, vertical error bars are at the 90\% statistical confidence level 
of the nominal model and horizontal bars indicate the radial binning 
size.
}
\label{fig:sys1}
\end{figure*}

\begin{figure*}
\includegraphics[width=0.82\textwidth, angle=270]{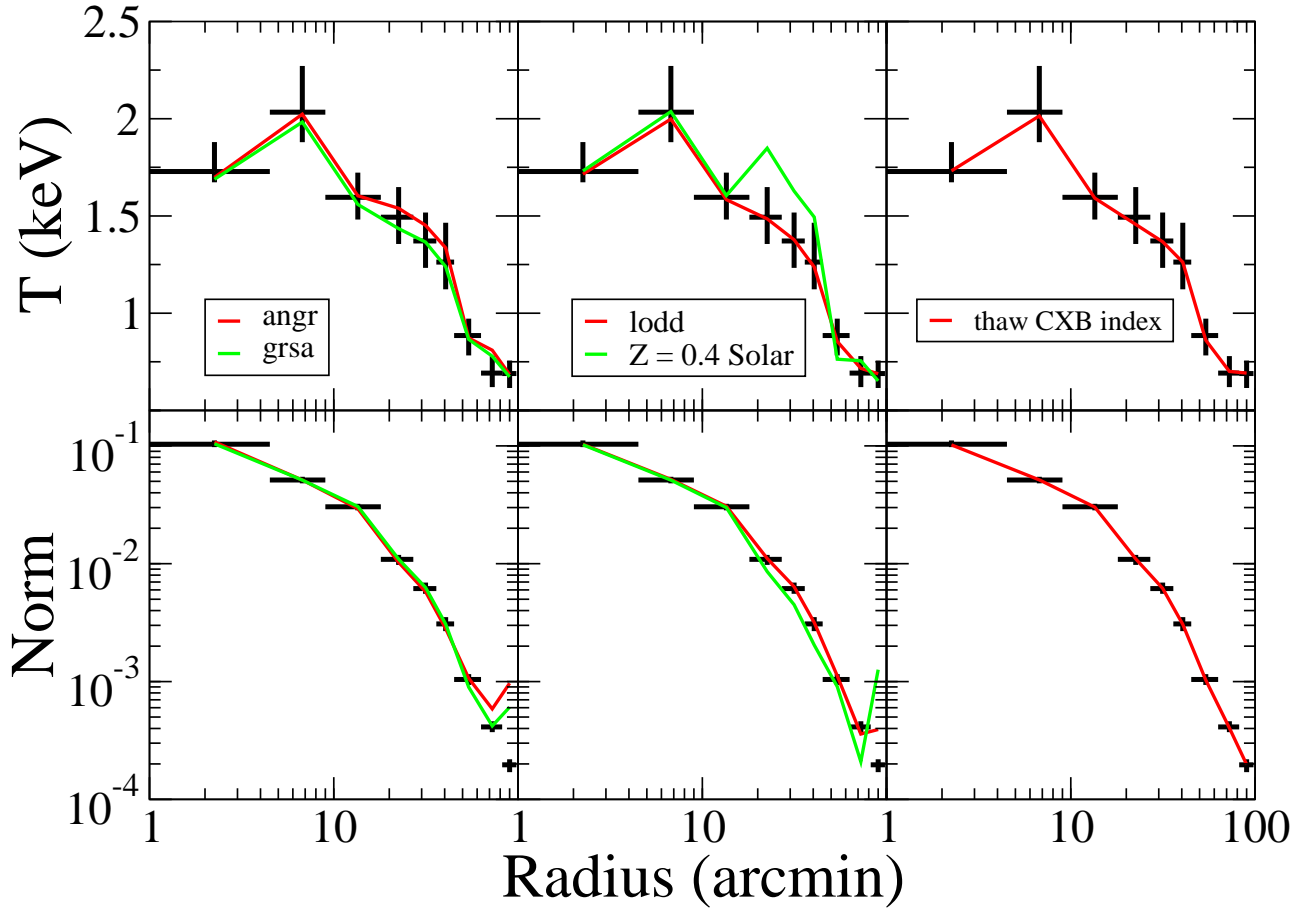}
\caption{
Left column: systematic uncertainties introduced by uncertainties in the solar 
abundance table.  
The upper panel shows the temperature profile of the 
nominal single temperature model using the {\tt aspl} table (black).
Model with the {\tt angr} ({\tt grsa}) table is shown in red (green).  
The lower panel 
shows the corresponding {\tt APEC} normalization per unit area.
Middle column:
similar to left column, but with the red line representing a model with 
the {\tt lodd} abundance table.
The green line shows the model using the {\tt aspl} table, but the metallicity
beyond 27\arcmin\ is fixed at 0.4 solar.
Right column: 
similar to left column, but with the red line representing model with 
thawed CXB photon index.
For all panels, vertical error bars are at the 90\% statistical 
confidence level of the nominal model and horizontal bars indicate the 
radial binning size.
}
\label{fig:sys2}
\end{figure*}

\begin{figure*}
\includegraphics[width=0.82\textwidth, angle=270]{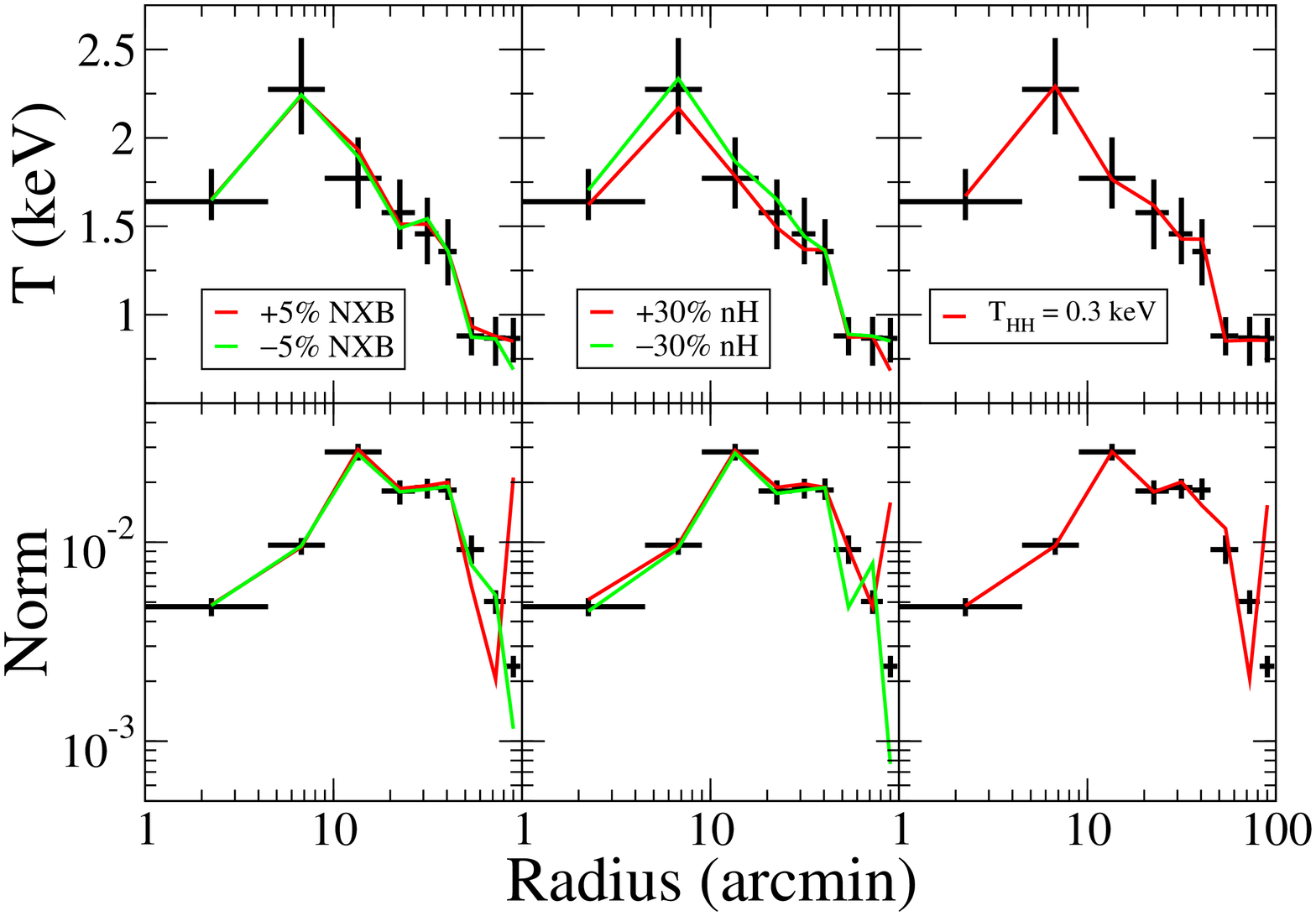}
\caption{
Similar to Figure~\ref{fig:sys1}, but for the deprojected spectral analysis
using the {\tt PROJCT} model.  
}
\label{fig:sys3}
\end{figure*}

\begin{figure*}
\includegraphics[width=0.82\textwidth, angle=270]{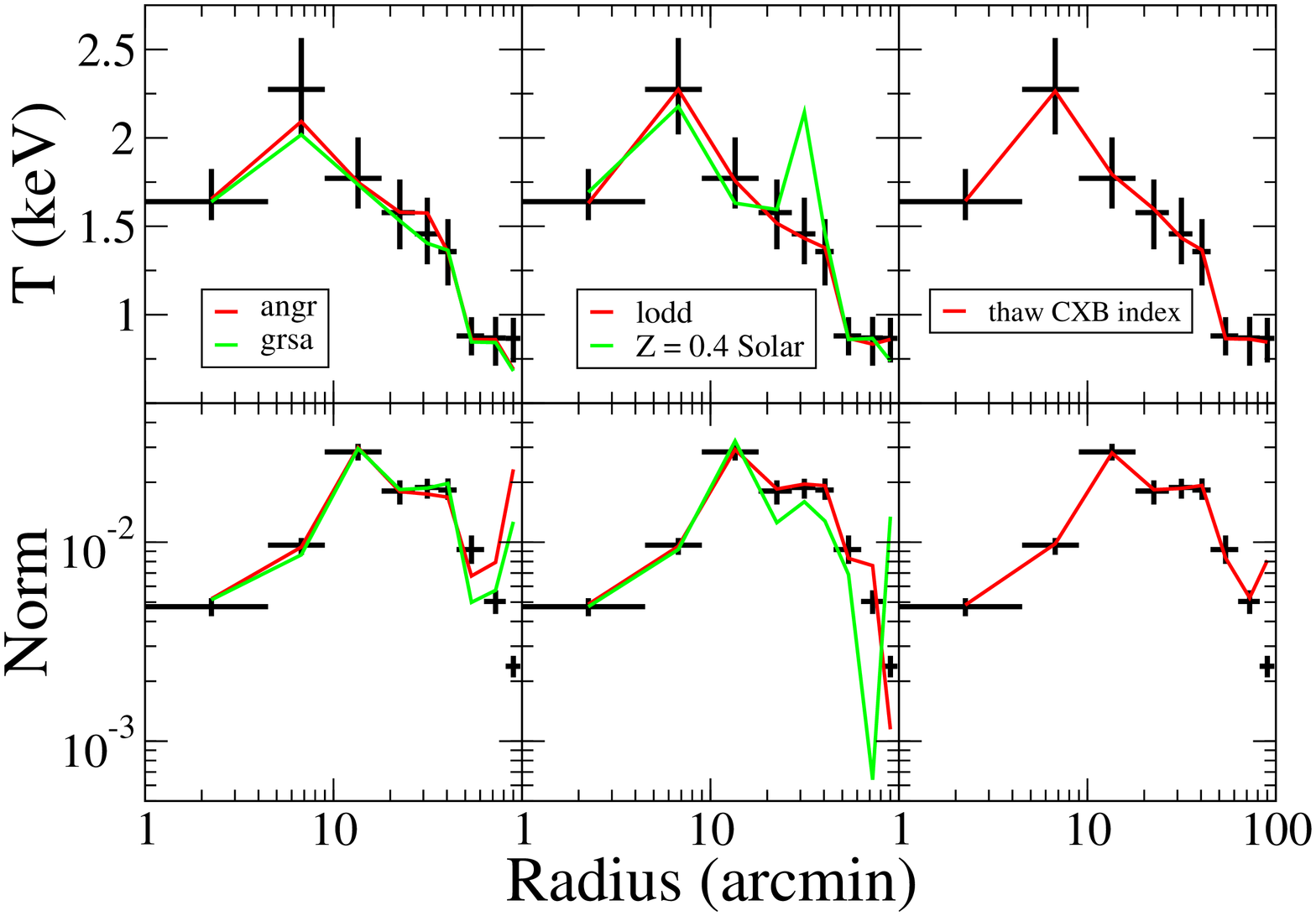}
\caption{
Similar to Figure~\ref{fig:sys2}, but for the deprojected spectral analysis
using the {\tt PROJCT} model. 
}
\vskip 1mm
\label{fig:sys4}
\end{figure*}

In Section~\ref{sec:obs}, we characterized the projected spectrum of the 
ICM component by the {\tt APEC} model with proper background modeling.  
To ensure that this characterization of the projected spectra is robust, we 
check it against systematic uncertainties in spectral modeling introduced 
by the background modeling, as well as the uncertainty in metallicity as 
outlined below.  All these systematic uncertainties are included in our 
data analysis.

The systematic uncertainty $\delta_{\rm sys}$ of each fitted parameter 
(temperature and gas normalization) is defined as the best-fit parameter 
minus the nominal best-fit parameter. We calculate the total upper ($u$) 
and lower ($l$) 90\% confidence errors by adding the systematic and 
statistic errors in quadrature:

\begin{equation}
\label{eq:delu}
\delta^2_{{\rm tot},u} = \sum_i \delta^2_{{\rm sys},i} + \delta^2_{{\rm 
stat},u} {\rm \,\,\, for \,\,\,} \delta_{{\rm sys},i} > 0
\end{equation}

and 

\begin{equation}
\label{eq:dell}
\delta^2_{{\rm tot},l} = \sum_i \delta^2_{{\rm sys},i} + \delta^2_{{\rm 
stat},l} {\rm \,\,\, for \,\,\,} \delta_{{\rm sys},i} < 0,
\end{equation}
where $i$ represents each of the systematic uncertainties described below.

\subsection{NXB uncertainties}
\label{sec:sys_NXB}

The NXB contributes from less than about 3(10)\% of the total
0.6--2.0 (0.5--7.0\,keV) emission at the center up to 
37(59)\% at the background field (EB).
We changed the background level by $\pm 
5\%$.  This generally introduces less than 3\% systematic uncertainties 
in projected temperature within $\sim$30\arcmin, but it is larger beyond that (left 
column in Figure~\ref{fig:sys1}).  
The projected temperature of some of the last data bins 
can be biased by $\sim$7\%, which is 
smaller than its statistical uncertainty.
Similarly, 
the projected gas normalization is biased by less than 4\% within 
$\sim$30\arcmin.  
It can be biased high by a factor of five in the outermost bin, which is 
significantly larger than its statistical uncertainty.  
The biases to the deprojected quantities are larger compared with 
the projected quantities, although the
statistical uncertainties are also larger.
The deprojected temperature within $\sim$30\arcmin\
can be biased up to 9\% (left column in Figure~\ref{fig:sys3}).  
The bias can be as large as 20\% 
at the outermost bin, which is comparable to its statistical uncertainty.  
The systematic uncertainties of the
deprojected normalizations within $\sim$45\arcmin\ are smaller than 
the statistical uncertainty.  For regions beyond that, the
deprojected normalizations can be biased by nearly an order of magnitude, 
highly subject to systematic uncertainties of the NXB.

\subsection{Galactic absorption uncertainties}
\label{sec:sys_nH}

We adopted the Galactic absorption values determined by \citet{WSB+13}, 
which are generally $\sim$25--30\% higher than those determined by 
\citet{KBH+05}.  We varied the Galactic values by $\pm30\%$ to address 
these systematic uncertainties.  The projected temperature is only biased 
slightly, up to $\sim$6\% near the center (middle column in 
Figure~\ref{fig:sys1}).  The biases in the projected normalizations are 
similarly small inside $\sim$63\arcmin.  Beyond that, the bias can be as 
large as a factor of four.  The systematic uncertainty in the 
deprojected temperature is also generally smaller than $\sim$5--6\%, 
except for the outermost bin that can be biased as large as $\sim$20\% 
(middle column in Figure~\ref{fig:sys3}).  The biases in deprojected 
normalizations are smaller then $\sim$5\% inside $\sim$45\arcmin, but 
again can be biased up to a factor of seven at the outermost bin.

\subsection{GXB uncertainties}
\label{sec:sys_GXB}

The temperature of the GXB hot halo was thawed in the spectral fitting, 
and its best-fit temperature is $T_{\rm HH} \approx 0.5$\,keV.  The high 
temperature may be due to the low Galactic latitude of Antlia.  We 
address the uncertainty of the hot halo temperature by fixing it at a 
value of 0.3\,keV, which is more typical for higher Galactic latitude.  
The low $T_{\rm HH}$ generally does not change the projected temperature 
by more than 3\% except at the outer region near 73\arcmin, where the 
bias is about 23\% (right column in Figure~\ref{fig:sys1}).  Similarly 
for the projected normalization, the biases are generally smaller than 
2\% except for the last data bin, where it can be largely biased by a 
factor of six.  For the deprojected temperature profile, the bias is at 
most 5\% everywhere (right column in Figure~\ref{fig:sys3}).  The 
systematic uncertainties of the deprojected normalizations are smaller 
than the statistical uncertainties inside $\sim$36\arcmin, while they are 
larger beyond that.  The bias can be as large as a factor of six in the 
outermost bin.

\subsection{Solar abundance table uncertainties}
\label{sec:sys_solartab}

Line emission is significant for the low temperature ICM in Antlia, 
and thus the uncertainty in the solar abundance table might introduce 
biases in the spectral analysis.  We used the {\tt aspl} solar abundance 
table for the nominal model, and assessed the systematic uncertainties 
by using the {\tt angr}, {\tt grsa}, and {\tt lodd} tables (left and 
middle columns in Figures~\ref{fig:sys2} and \ref{fig:sys4}).  The 
systematic biases in both the projected and deprojected temperatures are 
all smaller than or comparable to the statistical errors.  For the 
(de-)projected normalizations, the biases are generally less than a few (10) 
percent within $\sim$45\arcmin.  Beyond that, the biases increase, and 
can be as large as a factor of 5 (10) in the outermost bin.

\subsection{Metallicity uncertainties}
\label{sec:sys_abun}

The metallicity of the hot gas beyond $\sim$18\arcmin\ cannot be 
constrained with spectral fitting.  We fixed it to the lowest 
metallicity of 0.15\,$Z_{\odot}$ obtained with the inner region for our 
nominal model. The metallicity in the cluster outer regions can be as high 
as its central value (e.g., 0.3\,$Z_{\odot}$\footnote{Converted from 
the {\tt angr} to the {\tt grsa} abundance table.}
in Abell 399/401: \citealt{FTH+08};
0.3\,$Z_{\odot}$ in Perseus: \citealt{WUS+13}).
Therefore, we 
fixed the metallicity of Antlia to 0.4\,$Z_{\odot}$ beyond 
$\sim$18\arcmin, which is close to its central metallicity.
At temperatures below $\sim$1\,keV, the 
emission is dominated by line emission that is proportional to 
metallicity.  
This generally introduces a degeneracy between metallicity and gas 
density (because emission is also proportional to density squared).  
Setting the metallicity to 0.4\,$Z_{\odot}$ generally decreases 
the projected and deprojected gas normalizations by up to a factor of two and 
eight, respectively (middle  
columns in Figures~\ref{fig:sys2} and \ref{fig:sys4}). 
However, at the last data bin, both normalizations are biased high by a 
factor of six, which may be due to the effects coupled with the temperature and
background changes.
The high metallicity typically introduces a systematic uncertainty in 
temperature that is comparable to or larger than the statistical uncertainty.

\subsection{Unresolved CXB uncertainties}
\label{sec:sys_cxb}

The CXB was modeled with a fixed photon index of 1.4 \citep{KIM+02}. We 
assessed the systematic uncertainties by thawing the photon index (right 
columns in Figures~\ref{fig:sys2} and \ref{fig:sys4}). The best-fit 
photon index only changes slightly to 1.39 (1.37) for the (de-)projected 
spectral analysis, introducing systematic uncertainties in temperature 
that are smaller than the statistical uncertainties. The projected gas 
normalizations are only affected by $\lesssim 2\%$.  For the deprojected 
normalizations, the biases are at most up to $\lesssim 10\%$ and are smaller 
than the statistical errors inside $\sim$80\arcmin. Beyond that, it can 
be biased high by a factor of three, which is larger than the statistical 
uncertainty.

\end{document}